\documentclass[aoas]{imsart}

\RequirePackage{amsthm,amsmath,amsfonts,amssymb}
\RequirePackage[authoryear]{natbib}
\RequirePackage[colorlinks,citecolor=blue,urlcolor=blue]{hyperref}
\RequirePackage{graphicx}
\usepackage{changes}
\usepackage{latexsym, enumerate, subfigure}
\usepackage{amsbsy}
\usepackage{amsfonts}
\usepackage[mathcal]{euscript}
\usepackage{graphicx}
\usepackage{epsfig}
\usepackage{color}
\usepackage{enumerate,verbatim}
\usepackage{multirow}
\usepackage{xcolor}
\usepackage{url}
\usepackage[rflt]{floatflt}
\usepackage{tablefootnote}
\usepackage[symbol]{footmisc}
\usepackage{makecell}
\usepackage{nicematrix}

\def\bSig\mathbf{\Sigma}

\newcommand{\field}[1]{\mathbb{#1}}

\def\qed{\hfill$\diamondsuit$}

\def\argmin{\mathop{\mbox{argmin}}}

\newcommand{\be}{\begin{eqnarray}}
\newcommand{\ee}{\end{eqnarray}}
\newcommand{\ba}{\begin{eqnarray*}}
\newcommand{\ea}{\end{eqnarray*}}

\newcommand {\bfbeta} {\mbox{\boldmath $\beta$}}

\newcommand {\bfX} {{\bf X}}

\newcommand{\dkonv}{\stackrel{d}{\rightarrow}}

\startlocaldefs
\theoremstyle{plain}

\newtheorem{theorem}{Theorem}[section]

\theoremstyle{remark}

\endlocaldefs

\begin{document}

\begin{frontmatter}
\title{Decomposition of Longitudinal Disparities: an Application to the Fetal Growth-Singletons Study}
\runtitle{Decomposition of the Longitudinal Disparities}

\begin{aug}
\author[A, B]{\fnms{Sang Kyu}~\snm{Lee}\ead[label=e1]{sangkyu.lee@nih.gov}}\footnote[1]{Authors equally contributed},
\author[C]{\fnms{Seonjin}~\snm{Kim}\ead[label=e2]{kims20@miamioh.edu}}$^*$,
\author[D]{\fnms{Mi-Ok}~\snm{Kim}\ead[label=e3]{miok.kim@ucsf.edu}}
\author[E]{\fnms{Katherine L.}~\snm{Grantz}\ead[label=e4]{katherine.grantz@nih.gov}}
\author[B]{\fnms{Hyokyoung G.}~\snm{Hong}\ead[label=e5]{grace.hong@nih.gov}}
\address[A]{Department of Statistics and Probability,
Michigan State University}
\address[B]{Biostatistics Branch, Division of Cancer Epidemiology and Genetics,
National Cancer Institute\printead[presep={,\ }]{e1,e5}}
\address[C]{Department of Statistics, Miami University\printead[presep={,\ }]{e2}}
\address[D]{Department of Epidemiology and Biostatistics, University of California San Francisco\printead[presep={,\ }]{e3}}
\address[E]{Epidemiology Branch, Division of Population Health Research, Division of Intramural Research, Eunice Kennedy Shriver
National Institute of Child Health and Human Development\printead[presep={,\ }]{e4}}
\end{aug}

\begin{abstract}
Addressing health disparities among different demographic groups is a key challenge in public health. Despite many efforts, there is still a gap in understanding how these disparities unfold over time. Our paper focuses on this overlooked longitudinal aspect, which is crucial in both clinical and public health settings. In this paper, we introduce a longitudinal disparity decomposition method that decomposes disparities into three components: the explained disparity linked to differences in the exploratory variables' conditional distribution when the modifier distribution is identical between majority and minority groups, the explained disparity that emerges specifically from the unequal distribution of the modifier and its interaction with covariates, and the unexplained disparity.  The proposed method offers a dynamic alternative to the traditional Peters-Belson decomposition approach, tackling both the potential reduction in disparity if the covariate distributions of minority groups matched those of the majority group and the evolving nature of disparity over time. We apply the proposed approach to a fetal growth study to gain insights into disparities between different race/ethnicity groups in fetal developmental progress throughout the course of pregnancy. 
\end{abstract}


\end{frontmatter}

\section{Introduction}
\label{s:intro}

{
The assessment of fetal growth is an important part of routine prenatal care as an indication of fetal well-being \citep{ACOG2020macro, ACOG2021fetal}. Fetal growth restriction is associated with an increased risk of perinatal morbidity including hypoglycemia, hypothermia, and metabolic abnormalities as well as perinatal mortality \citep{nafday2017}. Fetal overgrowth, or macrosomia, is associated with an increased risk of birth trauma for both the parturient and the infant, and neonatal morbidity including hypoglycemia and hyperbilirubinemia \citep{nafday2017}.  The purpose of screening for fetal growth abnormalities is to implement increased monitoring such as antenatal testing to prevent a stillbirth and clinical interventions such as cesarean to prevent birth trauma in the case of suspected macrosomia. Emerging evidence suggests that incorporating growth velocity may improve clinical decisions that rely on estimated fetal weight \citep{sovio2015screening, grantz2018fetal}. This aspect of fetal monitoring is particularly critical for the early detection of any growth abnormalities, which could be indicative of more complex health issues. The determinants of fetal growth are not fully understood but a number of physiologic and pathologic factors have been identified including ethnic origin \citep{hanson2014early, GARDOSI2009}. Research shows fetal growth patterns can vary among racial and ethnic groups even in low-risk pregnancies with normal neonatal outcomes likely due to a complex interaction of genetic and environmental factors \citep{louis2015racial, workalemahu2018genetic, tekola2021admixture}. However, disparities in abnormal fetal growth also exist.  For example, in the U.S., infants of Black women are 1.5 times more likely to be small-for-gestational age than infants of White women \citep{centers2008}.  Given the recognition that fetal growth and birth size have lifelong implications for subsequent health and disease under the Developmental Origins of Health and Disease (DDHaD) Hypothesis, it is important to understand how these factors affect growth over time and their variations among different racial and ethnic groups is crucial for addressing this public health concern \citep{barker1990fetal, catalano2009perinatal, gluckman2010conceptual}.


 Disparity decomposition has been suggested as one approach to studying factors that influence health disparities in outcomes.
}
Numerous efforts have been made to investigate health disparities, specifically across race/ethnicity and sex \citep{naimi2016, jackson2018decomposition, jackson2021meaningful,lee2023quantile,park2023varying,Hong2024quantile}. However, traditional studies on disparity decomposition primarily rely on cross-sectional data, which falls short of addressing our objective of longitudinally examining fetal growth disparities across different racial and ethnic groups.

\citet{peters1941method} and \citet{belson1956technique} were the first to introduce the disparity decomposition method, which we will refer to as the {Peters-Belson (PB)} method. Later, \citet{oaxaca1973male} and \citet{blinder1973wage} developed a similar technique, further advancing this area of analysis. {The PB} approach involves separately fitting a regression model to data from the majority (advantaged) and minority (disadvantaged) groups. Following this, it uses the covariates from one group to generate a ``pseudo-outcome'' for the other. Building upon earlier work, they proposed a method to decompose the observed disparity, defined as the difference in the average outcomes between two groups, into two parts: the ``explained'' component, associated with differences in the values of explanatory variables, and the ``unexplained'' component. This latter component accounts for disparity arising from factors not captured by the explanatory variables in the model. Their approach allows for a more detailed decomposition of disparities, distinguishing between effects due to observable characteristics and those due to other, unmeasured factors.

Although PB method provides valuable insights into disparity analysis, it has limitations. The explained disparity in PB assumes that equalizing all covariate averages between minority and majority groups would pertain to the hypothetical reduction in disparities, which is unrealistic since interventions, if any, target a couple of specific covariates in practice. Moreover, projecting the effects of such interventions requires understanding how these targeted covariates interact with others, which the PB method does not handle well.
Although the PB framework has been successfully expanded to accommodate various response types  \citep{gastwirth1995biostatistical, sinclair2009using, fortin2011decomposition, graubard2005using, li2015extension,eberly2013extending,machado2005counterfactual},  to the best of our knowledge, there has been no research focused on examining disparities longitudinally.

In this paper, we extend the PB method to longitudinal settings, overcoming the constraints of the traditional PB approach that is tailored for cross-sectional data.  We treat the covariates that are potential targets of interventions as modifiers and introduce a new longitudinal disparity decomposition approach. The new approach captures the longitudinally evolving complex interactions with the modifiers using varying coefficient models. Our extension enables the exploration of disparity trends over time, thus providing a more comprehensive understanding. For instance, analyzing longitudinal patterns of fetal growth disparities can reveal when the divergence between minority and majority groups begins. This also allows for the estimation of a hypothetical growth trajectory for the minority group, assuming the same covariate characteristics as the majority group.

Additionally, we address a certain covariate that holds particular significance for complex interplay with other covariates. For clarity, these covariates are consistently referred to as `modifiers' throughout the text. This terminology is specifically chosen to aid in interpreting their potentially intricate interplay, especially as they develop over time. Our proposed approach provides a more comprehensive analysis of disparities by effectively handling these `modifiers', a capability absent in the conventional PB framework. 

The proposed methodological enhancement allows for a deeper and more comprehensive understanding of the dynamics involved in disparity analysis. This is accomplished by decomposing the disparity into three distinct components: one that aligns with the ``unexplained'' disparity in the traditional PB method, another that accounts for the ``explained'' disparity by all covariates except the modifier, when the modifier following the same hypothetical distribution is imposed on both the majority and minority groups and the other new component that captures the disparity attributed due to the modifier. Further, this new ``explained'' disparity considers the direct effect of the modifier and its indirect effect interacted with the covariates. Our method is capable of handling modifiers of both continuous and discrete types. To strengthen the statistical inference concerning the disparity, we introduce a bootstrap-based approach for generating simultaneous confidence bands, specifically designed for this time-varying disparity decomposition.

The remainder of this paper is organized as follows:
In Section 2, we provide an overview of the traditional PB method for cross-sectional data and present new methods for decomposing disparities in longitudinal settings, featuring two distinct approaches. In Section 3, we detail the methodology for estimating these disparities and conducting inference. In Section 4,  we introduce a tool designed to evaluate the influence of a modifier in scenarios specifically chosen by researchers. In Section 5, we demonstrate the practical application of our model using an empirical analysis of fetal growth data from the \textit{Eunice Kennedy Shriver} National Institute of Child Health and Human Development (NICHD) Fetal Growth Studies - Singletons \citep{Grewal2017}. Finally, in Section 6, we conclude with a discussion on the study's limitations and outline potential future directions.

\section{The Proposed Longitudinal Disparity Decomposition Methods}\label{sec:method}

\subsection{Longitudinal disparity decomposition (LDD)}\label{sec:PB}

First, we briefly introduce the traditional PB method for the cross-sectional data. Suppose the outcome of interest for the majority ($M$) and the minority ($m$) group can be modelled as following:
\begin{eqnarray*}
    Y^M &=& \textbf{X}^{M} \boldsymbol{\beta}^{M} + \epsilon^{M},\\
    Y^m &=& \textbf{X}^{m} \boldsymbol{\beta}^{m} + \epsilon^{m}\,,
\end{eqnarray*}
where $\textbf{X}^{M}$ and  $\textbf{X}^{m}$ 
are row vectors of exploratory variables,   $\boldsymbol{\beta}^{M}$ and $\boldsymbol{\beta}^{m}$ 
are column vectors of the respective regression coefficients, and $\epsilon^{M}$ and $\epsilon^{m}$ are  random variables with respective means equal to zero.
The overall disparity ($D$) between the majority and minority  groups is defined and decomposed as follows:
\begin{eqnarray}
    D &=& {E} (Y^{M}) - {E} (Y^{m})=  {E} \left\{ E(Y^{M}| \textbf{X}^{M})\right\} - E\left\{ E(Y^{m}| \textbf{X}^{m})\right\} \cr
    &=& {\  {E} \left\{ E(Y^{M}| \textbf{X}^{M})\right\} - E\left\{ E(Y^{m}| \textbf{X}^{M})\right\} +{E} \left\{ E(Y^{m}| \textbf{X}^{M})\right\} - E\left\{ E(Y^{M}| \textbf{X}^{m})\right\} }\cr
    &=&   \underbrace{{E} (\textbf{X}^{M}) ( \boldsymbol{\beta}^{M} - \boldsymbol{\beta}^{m} )}_{\textrm{\small{unexplained disparity}}} +  \underbrace{\left\{{E} (\textbf{X}^{M}) - {E} (\textbf{X}^{m}) \right\}\boldsymbol{\beta}^{m}}_{\textrm{\small{explained disparity}}}, \label{eq:DC} 
\end{eqnarray}
where $E\left\{ E(Y^{m}| \textbf{X}^{M})\right\}$, equivalently $E (\textbf{X}^{M}) \boldsymbol{\beta}^{m}$,  represents the expected value of a pseudo outcome, which would result if the minority group had consisted of members assuming the same exploratory values as the majority group. The first term is called ``unexplained disparity,'' as it represents the coefficient difference that is not accounted for by observed covariates or unmeasured factors.
The second term in  (\ref{eq:DC}) is termed ``explained disparity,'' as it captures the difference attributable to the observed covariates between groups.  More specifically, the explained disparity quantifies the potential reduction in disparity if the covariates' distributions of the minority group were to match those of the majority group. 

We note that the decomposition in (\ref{eq:DC}) diverges from the standard PB decomposition by adding and subtracting $E (\textbf{X}^{M}) \boldsymbol{\beta}^{m}$ instead of $E (\textbf{X}^{m}) \boldsymbol{\beta}^{M}$. The standard one defines the explained disparity as  $\left\{{E} (\textbf{X}^{m}) - {E} (\textbf{X}^{M}) \right\}\boldsymbol{\beta}^{M}$. This alternative formulation helps answer different kinds of research questions, such as understanding the impact of applying the majority group's exploratory variable structure ($E (\textbf{X}^{M})$) to the minority group. This alternative formulation would also be the natural choice for investigating the impact of modifiers by equalizing their distribution in the minority group to that in the majority group.

We now introduce an approach for longitudinal disparity decomposition, designed to assess decomposition at a specific time point,  $t$:
\begin{eqnarray}\label{eq:naive}
D(t)= \underbrace{E \{\mathbf{X}^{M}(t)\} \left\{ \boldsymbol{\beta}^{M}(t) - \boldsymbol{\beta}^{m}(t) \right\}}_\textrm{\small{unexplained disparity}}+\underbrace{\left[E \{\mathbf{X}^{M}(t)\} - E \{\mathbf{X}^{m}(t)\} \right]\boldsymbol{\beta}^{m}(t)}_\textrm{\small{explained disparity}}.
\end{eqnarray}

This method naively extends the PB method to longitudinal analysis. We refer to the disparity decomposition approach presented in (\ref{eq:naive}) as the longitudinal disparity decomposition (LDD).
The estimation of the overall disparity \(D(t)\), along with the estimations of the explained and unexplained disparities, will be discussed in Section 3.

\subsection{Longitudinal disparity decomposition with modifier (mLDD)}\label{sec:longdisparity}

Although the approach proposed in Section 2.1, namely LDD, can measure the explained disparity by the covariates included in the model, it collectively aggregates the influences of all these covariates. This may not be ideal if our interest lies in a specific modifier, say \(Z\), that is amenable to intervention. In such cases, it becomes important to consider the effect of \(Z\) and its interaction with the remaining covariates, denoted by \(\mathbf{X}\).  For example, enhancing education, represented by $Z$, might directly reduce disparities in health outcomes.  Moreover, such educational improvements could interact with other variables,   like socioeconomic or physiological status, potentially contributing further to disparity reduction. This section aims to understand the disparity arising from the direct effect of \(Z\) as well as the indirect effects stemming from \(Z\)'s interactions with other covariates, denoted as \(\mathbf{X}\).

First, we introduce a time-varying coefficient longitudinal model in which the coefficients are functions of both the measurement time \(t\) and the modifier \(Z\):
\begin{eqnarray*}
Y(t) &=& \beta_0(t, Z) + \beta_1(t, Z)X_1(t) + \cdots + \beta_{p}(t, Z)X_p(t) + \sigma(t, Z)\epsilon \\
&=& \mathbf{X}(t)\boldsymbol{\beta}(t, Z) + \sigma(t, Z)\epsilon,
\end{eqnarray*}
where \(\mathbf{X}(t) = (1, X_1(t), \dots, X_p(t))\), \(\boldsymbol{\beta}(t, Z) = (\beta_0(t, Z), \dots, \beta_p(t, Z))^T\) and \(\epsilon\) is an error term with \(E(\epsilon) = 0\) and \(E(\epsilon^2) = 1\). Here, \(\beta_0(t, Z)\) represents the influence of \(Z\) on \(Y(t)\) across time \(t\), and \(\beta_1(t, Z), \dots, \beta_p(t, Z)\) denote the influences of the covariates \(X_1(t), \dots, X_p(t)\) on \(Y(t)\), conditioned on \(Z\). Since the coefficients of \(\mathbf{X}\) are functions of  \(Z\), the varying coefficient model can be viewed as a regression model with flexible forms of interaction terms between \(Z\) and \(\mathbf{X}\). Due to the nonparametric specification of the coefficients, the modifier, although can be multivariate in theory, may often be a univariate variable. In this paper, we focus on a univariate modifier, \(Z\).

Let  \(Z^M\) and \(Z^m\) represent the modifiers for the majority and minority groups respectively. We define the longitudinal disparity at time \(t\) between the majority (\(M\)) and minority (\(m\)) groups, conditioned on \(Z^M\) and \(Z^m\), as:
\begin{eqnarray}
     D(t) = {E} \{Y^{M}(t)\} - {E} \{Y^{m}(t)\}=E\{D(t|Z^M,Z^m)\}, \label{eq:1}
\end{eqnarray}
where 
\begin{eqnarray*}
D(t|Z^M,Z^m)&=&E\{Y^M(t)|Z^M\}-E\{Y^m(t)|Z^m\}\cr
&=&E\{\bfX^M(t)|Z^M\}\bfbeta^M(t,Z^M)- E\{\bfX^m(t)|Z^m\}\bfbeta^m(t,Z^m).
\end{eqnarray*}
This framework enables us to investigate the impact of the modifier $Z$. $Z$ not only alters the mean of $\mathbf{X}(t)$ (expressed as $E\{\mathbf{X}(t)|Z\}$), but also affects how $\mathbf{X}(t)$ influences $Y(t)$ through their coefficients $\bfbeta(t,Z)$. In order to see this clearly, we further decompose (\ref{eq:1})  by adding  $E[E\{\bfX^M(t)|Z^M\}\bfbeta^m(t,Z^M)]$ and \newline $E[E\{\bfX^m(t)|Z^M\}\bfbeta^m(t,Z^M)]$ to and subtracting the same quantities from (\ref{eq:1}) as follows:
\begin{eqnarray}
 D(t)&=& D_1(t) + D_2(t) +D_3(t) \cr 
&:=& \underbrace{E[E\{\bfX^M(t)|Z^M\}\{\bfbeta^M(t,Z^M)-\bfbeta^m(t,Z^M)\}]}_\textrm{\small{unexplained disparity}}\label{eq:2}\\
&+&\underbrace{E([E\{\bfX^M(t)|Z^M\}-E\{\bfX^m(t)|Z^M\}]\bfbeta^m(t,Z^M))}_\textrm{\small{explained disparity due to $\bfX$ when $Z^M$ imposed}}\label{eq:3}\\
&+&\underbrace{E[E\{\bfX^m(t)|Z^M\}\bfbeta^m(t,Z^M)-E\{\bfX^m(t)|Z^m\}\bfbeta^m(t,Z^m)]}_\textrm{\small{explained disparity due to the modifier $Z$}}.\label{eq:4}
\end{eqnarray}
 The first term, \(D_1(t)\) in (\ref{eq:2}), captures disparity arising from differences in the coefficients and represents the unexplained disparity.  Conversely, \(D_2(t)\) captures the explained disparity, which stems from differences in the covariates \(\mathbf{X}\) between the majority and minority groups, under a hypothetical scenario where the modifiers in both the majority and the minority group are arbitrarily set to follow the modifier distribution in the majority group. The measure of \(D_2(t)\) involves averaging the conditional expectations of \(\mathbf{X}\) over all possible values of \(Z^M\) across the entire distribution of \(Z^M\). For example, when the education level distribution from the majority group is imposed on both groups, \(D_2(t)\) quantifies the portion of the overall disparity associated with the differences in \(\mathbf{X}^M(t)\) and \(\mathbf{X}^m(t)\).
Finally, \(D_3(t)\) measures the portion of the explained disparity due to the distributional difference between \(Z^M\) and \(Z^m\), which includes its direct and indirect effect via \(\mathbf{X}\). The indirect effect is determined based on the interaction between \(\mathbf{X}^m(t)\) and the differing values of \(Z^M\) and \(Z^m\). 
Thus, the measure \(D_3(t)\) is utilized to assess the potential reduction in disparity for the minority group when the distributions of \(Z^M\) and \(Z^m\) are made equivalent. Specifically, the value of \(D_3(t)\) is zero if the distributions of \(Z^M\) and \(Z^m\) are identical. This unique characteristic of \(D_3(t)\) serves as a metric for understanding the impact of equalizing the disparate distributions of the modifier on the overall disparity. Henceforth this method will be referred to as the longitudinal disparity decomposition with modifier (mLDD).

Compared to LDD, mLDD further addresses the role of the modifier \(Z\) in explaining the disparity. This approach is particularly valuable in contexts where the overall effect of \(Z\) is of interest, encompassing its entire distribution rather than focusing on specific values.

\section{Estimation and inference}\label{sec:Est}

In this section, we introduce a method for estimating the longitudinal disparity decomposition within the framework of mLDD. 
We omit the specific estimation methods for   LDD  since LDD can be regarded as a special case of mLDD with the following representation of $Y(t)$: 
\begin{eqnarray}\label{eq:mLDD}
Y(t)=\beta_0(t)+\beta_1(t)X_1(t)+\cdots+\beta_{p}(t)X_p(t)+\beta_{p+1}(t)Z(t)+\sigma(t)\epsilon.
\end{eqnarray}

We let $Y_i(t)$, $Z_i$, and $\bfX_i(t)=\{1, X_{i1}(t),\dots,X_{ip}(t)\}$ represent the respective variables for subject $i$ measured at $t$ and 
$t_{ij}$ represents the measurement time within the time  period ${\cal T}$. Then random samples from the majority and minority groups are denoted as 
 $\{ \bfX_{i}^M(t_{ij}^M
 ),Y_{i}^M(t_{ij}^M),t_{ij}^M,Z_i^M\}$, $i=1,\ldots,n^M, j=1,\dots,n_i$,
 and  
 $\{ \bfX_{i}^m(t_{ij}^m),Y_{i}^m(t_{ij}^m),t_{ik}^m,Z_i^m\}$, $i=1,\ldots,n^m,  j=1,\dots,n_i$ respectively. 

Assuming independence between the majority and the minority group variables, 
by the definition of the longitudinal  disparity in (\ref{eq:1}), we have 
\begin{eqnarray}\label{eq:estimate}
    D(t)=\int\int  D(t|z^M,z^m)f_{Z^M}(z^M)f_{Zm}(z^m) dz^Mdz^m.
\end{eqnarray}
The empirical version of   (\ref{eq:estimate}) is
\begin{eqnarray*}
&&\frac{1}{n^Mn^m}\sum_{i=1}^{n^M}\sum_{i'=1}^{n^m}D(t|z_i^M,z_{i'}^m)\\
&=&\frac{1}{n^Mn^m}\sum_{i=1}^{n^M}\sum_{i'=1}^{n^m} E\{\bfX^M(t)|z_{i}^M\}\{\bfbeta^M(t,z_{i}^M)-\bfbeta^m(t,z_{i}^M)\} \\
&+&\frac{1}{n^Mn^m}\sum_{i=1}^{n^M}\sum_{i'=1}^{n^m}[E\{\bfX^M(t)|z_{i}^M\}-E\{\bfX^m(t)|z_{i}^M\}]\bfbeta^m(t,z_{i}^M)\\
&+& \frac{1}{n^Mn^m}\sum_{i=1}^{n^M}\sum_{i'=1}^{n^m}\left[ E\{\bfX^m(t)|z_{i}^M\}\bfbeta^m(t,z_{i}^M)-E\{\bfX^m(t)|z_{i'}^m\}\bfbeta^m(t,z_{i'}^m)\right],
\end{eqnarray*}
where $\bfbeta(t,z)$ and $E\{\bfX(t)|z\}$ for the majority and minority groups should be estimated at different $t$'s and $z$'s.

\subsection{Estimation}\label{sec:est}

The estimation method for the varying coefficient $\bfbeta(t,z)$ differs based on whether the modifier $Z$ is continuous or discrete. If the modifier $Z$ is a continuous random variable, we can obtain the estimator of $\bfbeta(t,z)$, denoted as  $\hat{\bfbeta}(t,z)$,  by   minimizing
\begin{eqnarray}\label{eq:localC}
\sum_{i=1}^n\sum_{j=1}^{n_i}\left\{Y_{i}(t_{ij})-\bfX_{i}(t_{ij})\bfbeta\right\}^2K\left(\frac{t_{ij}-t}{b_1}\right)K\left(\frac{Z_{i}-z}{b_2}\right),
\end{eqnarray}
where $K(\cdot)$ represents a kernel function and 
 $b_1$ and $b_2$ are the bandwidths. Here, we opt for the local constant estimator to estimate $\hat{\bfbeta}(t,z)$ over alternative methods such as local polynomial or spline-based approaches. We make this choice because the local constant estimator offers both simplicity and ease of interpretation. Moreover, given the context of our specific example in Section \ref{sec:app}, we do not anticipate rapid changes in the coefficients over time, making the local constant estimator a suitable choice.

If the modifier $Z$ is a discrete random variable, (\ref{eq:localC}) can be modified as
\begin{eqnarray}\label{eq:localD}
\sum_{i=1}^n\sum_{j=1}^{n_i}\left\{Y_{i}(t_{ij})-\bfX_{i}(t_{ij})\bfbeta\right\}^2K\left(\frac{t_{ij}-t}{b_1}\right)I(Z_i=z),
\end{eqnarray}
where $I(\cdot)$ represents the indicator function; specifically, $I(Z_i=z)$ selects a subsample of subjects who share the same level of the discrete random variable $Z$.

The following Theorem \ref{thm:1} establishes that the local constant estimator $\hat{\bfbeta}(t,z)$ is a consistent estimator for $\bfbeta(t,z)$. Additionally, it shows that the asymptotic distribution of $\hat{\bfbeta}(t,z)$ is normally distributed. 

\begin{theorem}\label{thm:1}

(a) Suppose that $Z$ is a continuous random variable and let $t$ and $z$  be interior points in their respective ranges. Let $N=\sum_{i=1}^nn_i$. Under assumptions (A1)--(A4) in Appendix, as $Nb_1b_2\to \infty$,  $Nb_1b_2(b_1^8+b_2^8)\to 0$ and $N^{-1}(1/\sqrt{Nb_1b_2}+b_1b_2)\sum_{i=1}^nn_i^2\to \infty$, 
 we have
$$
\sqrt{Nb_1b_2}\left\{\hat{\bfbeta}(t,z) -\bfbeta(t,z)-\frac{\rho(t,z)\phi_K}{2}\right\}\dkonv N\left\{0,\frac{\sigma^2(t,z)\psi_K^2}{f_{T,Z}(t,z)} \Gamma^{-1}_X(t)\right\},
$$
where  $f_{T,Z}(\cdot,\cdot)$ are the density functions of $t_{ij}$ and $Z_i$, $\psi_K=\int K^2(v)dv$,$\phi_K=\int v^2K(v)dv$, $\Gamma_X(t)=E\{\bfX(t)^T\bfX(t)\}$, and $
\rho(t,z)=b_1^2\{\bfbeta''_t(t,z)+2\bfbeta'_t(t,z)f_t'(t,z)/f_{T,Z}(t,z)\}+\newline b_2^2\{\bfbeta''_z(t,z)+2\bfbeta'_z(t,z)f_z'(t,z)/f_{T,Z}(t,z)\}
$. The derivatives are defined in the Appendix. 
\vspace{0.1in} 

(b) Suppose that $Z$ is a discrete random variable and let $t$ be an interior point within its range.  Under assumptions (A1)--(A4) in Appendix and $q_z=P(Z_i=z)>0$, as $Nb_1\to \infty$,  $ Nb_1^9\to 0$, and $N^{-1}(1/\sqrt{Nb_1}+b_1)\sum_{i=1}^nn_i^2\to \infty$ 
 we have
\begin{eqnarray*}
    &&\sqrt{Nq_zb_1}\left[\hat{\bfbeta}(t,z) -\bfbeta(t,z)-b_1^2\phi_K\left\{\frac{\bfbeta''_t(t,z)}{2}+\frac{\bfbeta'_t(t,z)f'_T(t)}{f_T(t)}\right\}\right]\cr
    &&\dkonv N\left\{0,\frac{\sigma^2(t,z)\psi_K^2}{f_T(t)} \Gamma^{-1}_X(t)\right\}.
\end{eqnarray*}
\end{theorem}
The conditions $N^{-1}(1/\sqrt{Nb_1b_2}+b_1b_2)\sum_{i=1}^nn_i^2\to \infty$ and  $N^{-1}(1/\sqrt{Nb_1}+b_1)\sum_{i=1}^nn_i^2\newline\to \infty$ are hold for both sparse longitudinal data ($\sup_i n_i\leq M$ for a constant $M$) and dense longitudinal data $(\sup_i n_i \to \infty)$ if $\sup_i n_i \{b_1b_2+(nb_1b_2)^{-1}\}\to 0$ and $\sup_i n_i \{b_1b+(nb_1b)^{-1}\}\to 0$, respectively. We note that most longitudinal data in practice have sparse or moderately dense observations. Furthermore, the asymptotic results in Theorem 1 are obtained without imposing any dependence structure on the error process. Instead, we assume a rather loose bound on $\bfbeta(t,z)$ in (A3) in the supplementary material.

Next, our objective is to estimate the conditional mean of covariates $X_r(t)$ given $Z=z$, denoted as $E\{X_r(t)|z\}$.
In case where $Z$ is continuous,  we employ a local constant estimator, $\hat\mu_{X_r}(t,z)$, for $r=1,\dots,p$, to approximate  $E\{X_r(t)|Z=z\}$.  The estimator is defined as follows:
\begin{eqnarray*}
\hat\mu_{X_r}(t,z)=
\argmin_{\mu_{X_r}}\sum_{i=1}^n\sum_{j=1}^{n_i}\left\{X_{i}(t_{ij})-\mu_{X_r}\right\}^2K\left(\frac{t_{ij}-t}{b_{r1}}\right)K\left(\frac{Z_{i}-z}{b_{r2}}\right).
\end{eqnarray*}
Alternatively, when $Z$ is a discrete variable,  the estimator $\hat\mu_{X_r}(t,z)$ takes the following form: 
\begin{eqnarray*}
\hat\mu_{X_r}(t,z)=
\argmin_{\mu_{X_r}}\sum_{i=1}^n\sum_{j=1}^{n_i}\left\{X_{i}(t_{ij})-\mu_{X_r}\right\}^2K\left(\frac{t_{ij}-t}{b_{r1}}\right)I(Z_i=z).
\end{eqnarray*}
The consistency and asymptotic distribution of $\hat\mu_{X_K}(t,z)$ can be established in a manner analogous to Theorem \ref{thm:1}.

 \subsection{Implementation}

To select appropriate bandwidths $b_1$ and $b_2$ in (\ref{eq:localC}), we utilize the leave-one subject-out cross-validation method. Specifically, we estimate $\hat{\bfbeta}^{(-i)}(t_{ij},Z_i)$ by excluding  the $i$th subject and  minimizing the following objective function: 
\begin{eqnarray}\label{eq:bandwidth}
\sum_{i'=1}^n\sum_{j=1}^{n_{i'}}\left\{Y_{i'}(t_{i'j})-\bfX_{i'}(t_{i'j})\bfbeta\right\}^2I(i' \neq i) K\left(\frac{t_{i'j}-t_{ij}}{b
_1}\right)K\left(\frac{Z_{i'}-Z_i}{b_2}\right).
\end{eqnarray}
The optimal bandwidths are determined by minimizing the following:
$$
(b_1^*,b_2^*)=\argmin_{b_1,b_2}\sum_{i=1}^n\sum_{j=1}^{n_i} [Y_{i}(t_{ij})-\bfX_{i}(t_{ij})\hat{\bfbeta}^{(-i)}(t_{ij},Z_i)]^2.
$$
When the modifier is discrete, only $b_1$ 
  is required. In this case, we  substitute the term 
$K\{(Z_{i'}-Z_i)/b_2\}$  in (\ref{eq:bandwidth}) with $I(Z_{i'}=Z_i)$. Subsequently, the leave-one subject-out cross-validation method is employed for bandwidth selection. The bandwidths $b_{r1}$ and $b_{r2}$ in the estimation of $E\{X_r(t)|z\}$ are also selected in the same fashion as $b_1$ and $b_2$ in estimation of $\bfbeta(t,z)$.

Finally, for a given time $t \in {\cal T}$, we estimate the unexplained disparity as in (\ref{eq:2}), the explained disparity due to  covariates $\bfX$ when $Z^M$ imposed as in (\ref{eq:3}),  the disparity explained by the modifier as in (\ref{eq:4}), and the  overall disparity as in (\ref{eq:1}) by summing up them as follows: 
\begin{eqnarray*}
\hat{D}(t)&=&\hat{D}_1(t) + \hat{D}_2(t) +\hat{D}_3(t)\\
&:=&\frac{1}{n^M}\sum_{i=1}^{n^M}\hat{E}\{\bfX^M(t)|z_{i}^M\}\{\hat{\bfbeta}^M(t,z_{i}^M)-\hat{\bfbeta}^m(t,z_{i}^M)\}\\
&+&\frac{1}{n^M}\sum_{i=1}^{n^M}[\hat{E}\{\bfX^M(t)|z_{i}^M\}-\hat{E}\{\bfX^m(t)|z_{i}^M\}]\hat{\bfbeta}^m(t,z_{i}^M)\\
&+&\frac{1}{n^M}\sum_{i=1}^{n^M}\hat{E}\{\bfX^m(t)|z_{i}^M\}\hat{\bfbeta}^m(t,z_{i}^M)-\frac{1}{n^m}\sum_{i'=1}^{n^m}\hat{E}\{\bfX^m(t)|z_{i'}^m\}\hat{\bfbeta}^m(t,z_{i'}^m).
\end{eqnarray*}
Since $\hat{E}\{\bfX(t)|z\}$ and $\hat{\bfbeta}(t,z)$ are consistent estimators, it follows that $\hat{D}_1(t)$, $\hat{D}_2(t)$, $\hat{D}_3(t)$, and consequently $\hat{D}(t)$, are also consistent.

\subsection{Inference on the decomposed disparities}\label{sec:boot}

To draw statistical inferences regarding the longitudinal disparity  $D(t)$ and its decomposition $D_1(t)$, $D_2(t)$  and $D_3(t)$ in mLDD, we employ a simultaneous confidence band (SCB). This allows us to assess whether these quantities deviate significantly from zero over the range  $t\in {\cal T}$. 
A $(1-\alpha)100\%$ SCB for  $D(t)$ is defined as follows:
$$
P\left[\sup_{t \in {\cal T}} \left\{\frac{|\hat{D}(t)- D(t)|}{\hat{s}(t)} \right\}\leq Q_\alpha   \right]=1-\alpha,
$$
where $\hat{s}(t)$ represents the standard error of $\hat{D}(t)$. SCBs for  $D_1(t)$, $D_2(t)$ and $D_3(t)$ can be constructed in a  similar manner. 
Specifically, we utilize a bootstrap SCB framework, as proposed by \citet{kim2021predictive}. The details of this approach are elaborated below.

\begin{enumerate}
\item Obtain $\hat{D}(t)$ for $t\in {\cal T}$.
\item Generate $B$ bootstrap samples of size $n^M$ of the majority group by sampling subjects in 
the majority group with replacement. Generate $B$ bootstrap samples of size $n^m$ of the minority group in the same manner. 
\item Obtain $\hat{D}^{(b)}(t)$, for  $b=1,\dots,B$ and $t\in {\cal T}$,  with each bootstrap sample. 
\item At given $t\in {\cal T}$, obtain $\hat{s}(t)$ by computing the sample standard deviation of $\hat{D}^{(b)}(t)$ for $b=1,\dots,B$.
\item For each bootstrap estimate, compute the supremum of absolute standardized deviance from the estimate $\hat{D}(t)$ with the  estimate based on the original dataset:
\begin{eqnarray}\label{eq:sup}
Q^{(b)}:=\sup_{t \in {\cal T}} \left\{\frac{|\hat{D}^{(b)}(t)- \hat{D}(t)|}{\hat{s}(t)} \right\}.
\end{eqnarray}
\item Construct a $(1-\alpha)100\%$ SCB for $D(t)$ with 
$$
(\hat{D}(t)-\hat{s}(t)\hat{Q}_\alpha,\hat{D}(t)+\hat{s}(t)\hat{Q}_\alpha),  \quad   \mbox{all  }  t \in {\cal T},
$$
where $\hat{Q}_\alpha$ is  a $(1-\alpha)100\%$ percentile of $Q^{(b)}$.  
The proposed SCB contains $(1-\alpha)100\%$ of bootstrap estimates over  $t \in {\cal T}$.
\end{enumerate}

The R package, \texttt{vcPB}, has been developed to perform longitudinal disparity analysis and is available at \url{https://sangkyustat.github.io/vcPB/} or on CRAN.

\section{Conditional longitudinal disparity decomposition with a modifier (cmLDD)}\label{sec:cond}

The LDD and mLDD, discussed in Section \ref{sec:method}, examine the marginal disparity $D(t)=E\{Y^M(t)\}-E\{Y^m(t)\}$ by integrating the distribution of covariates including the modifier $Z$. Especially, the proposed mLDD offers a comprehensive approach for evaluating the impact of modifiers on disparity. However, practical scenarios often demand a more nuanced exploration of disparities, especially when examining differences between majority and minority groups at specific levels of a modifier. For example, when analyzing health outcome disparities between well-educated individuals in the majority group and under-educated individuals in the minority group or vice versa, the mLDD method may prove inadequate. To tackle these challenges, we present a new approach called the conditional longitudinal disparity decomposition with a modifier (cmLDD).

The cmLDD approach allows for a more targeted analysis of disparity, particularly in cases where a key modifier, such as education, significantly contributes to the observed disparities. To implement this approach, we consider the following model:
\begin{eqnarray}
D(t|z^M,z^m)&=&E\{Y^M(t)|z^M\}-E\{Y^m(t)|z^m\}\cr
&=&E\{\bfX^M(t)|z^M\}\bfbeta^M(t,z^M)- E\{\bfX^m(t)|z^m\}\bfbeta^m(t,z^m),\label{eq:cond}
\end{eqnarray}
where $z^M$ and $z^m$ are specific values of the modifier of interest.

We further decompose (\ref{eq:cond}) similarly to the method described previously in  (\ref{eq:2}) - (\ref{eq:4}) :
\begin{eqnarray}\label{eq:cmLDD}
 D(t|z^M,z^m)&=&  D_1(t|z^M)+D_2(t|z^M)+D_3(t|z^M,z^m)\\
 &:=&\underbrace{E\{\bfX^M(t)|z^M\}\{\bfbeta^M(t,z^M)-\bfbeta^m(t,z^M)\}}_\textrm{\small{unexplained disparity}}\cr
&+&\underbrace{[E\{\bfX^M(t)|z^M\}-E\{\bfX^m(t)|z^M\}]\bfbeta^m(t,z^M)}_\textrm{\small{explained disparity due to $\bfX$ when $z^M$ imposed to both $Z^M$ and $Z^m$}}\cr
&+&\underbrace{E\{\bfX^m(t)|z^M\}\bfbeta^m(t,z^M)-E\{\bfX^m(t)|z^m\}\bfbeta^m(t,z^m)}_\textrm{\small{explained disparity due to the difference between $z^M$ and $z^m$}}.\label{eq:cmLDDdecom}
\end{eqnarray}

In contrast to mLDD, which examines disparity by considering $Z$ in contexts focused on the distributional effect of $Z$ rather than specific values, the cmLDD method is specifically designed to address longitudinal disparities based on distinct values of modifiers for the majority and minority groups. More specifically, cmLDD evaluates disparity by analyzing the differences in conditional expectations of the outcome for certain values of $Z$ (such as high or low education levels) across majority and minority groups.

This tailored approach becomes particularly crucial in understanding potential shifts in disparity under different scenarios involving $z^M$ and $z^m$. These scenarios may encompass comparisons between the majority with a low education level and the minority with a high level, or vice versa. The cmLDD methodology provides a nuanced perspective on disparity, focusing on specific values of the modifier and offering valuable insights into how these variations impact the observed disparities between majority and minority groups over time.

Due to the law of total expectation $D(t)=E\{D(t|Z^M,Z^m)\}$,  the decomposition in cmLDD in (\ref{eq:cmLDD}) can be estimated concurrently with the estimation of their counterparts in mLDD. The statistical inference in Section \ref{sec:boot} is also easily applicable to cmLDD.

\section{Applications to fetal growth decomposition}\label{sec:app}

\subsection{NICHD Fetal Growth Studies - Singletons}

We utilize data from the NICHD Fetal Growth Studies, a prospective cohort study of singleton births aiming to establish standard fetal growth benchmarks by gestational age in the U.S. population \citep{louis2015racial, Grewal2017}. Participants are low-risk pregnant women from four racial and ethnic groups recruited at 12 U.S. clinical centers from 2009 to 2013. {According to the U.S. Natality Data \citep{osterman2023births}, the distribution of racial and ethnic groups is as follows: Non-Hispanic White (NHW) at 52\%, Non-Hispanic Black (NHB) at 14\%, Hispanic (HIS) at 24\%, and Asian or Pacific Islander (API) at 6\%. To align with these statistics,  we set NHW as the majority group ($M$), and NHB, HIS, and API as the minority groups ($m$).}

Eligibility criteria for the study encompass maternal age ranging from 18 to 40 years, {non-obese}, a pre-pregnancy BMI between 19.0 and 29.9, and a viable singleton pregnancy confirmed between 8 weeks 0 days and 13 weeks 6 days. The gestational age has to be consistent with the last menstrual period, as verified by a screening sonogram within a specific range. The study excludes women with previous pregnancy complications, chronic illnesses, medically assisted pregnancies, smoking, drug use, excessive alcohol consumption, preterm delivery, gestational diabetes, and neonatal problems like birth defects and mortality.

After an initial sonogram between 10 to 13 weeks of gestation, participants are randomly assigned to one of four follow-up schedules. These schedules include five more sonograms at gestational weeks 16-22, 24-29, 30-33, 34-37, and 38-41. We assess fetal growth using the Hadlock formula \citep{hadlock1985estimation}, which incorporates head circumference (HC), abdominal circumference (AC), and femur length (FL) measurements.

Alongside the fetal measurements, the study gathers demographic and medical history from the mothers. These include employment status, age, number of previous births (parity), marital status, insurance status, income, and level of education.

The original dataset consisted of 15,882 fetal growth measurements from 2,802 women. After excluding records related to high-risk pregnancies and missing data, the refined dataset included 7,543 fetal growth measurements from 1,464 pregnant women.

In our analysis, we considered the following maternal characteristics as covariates, adapted from \cite{GARDOSI2009}: age (in years), height (in cm), weight  (in kg) measured at enrollment,  past parity status (coded as `1' for multiparas, or having had a previous birth, and `0' for nulliparas, or no previous births), employment or student status (`1' for full-time employed or student, and `0' for others), marital status (`1' for married or living with a partner, and `0' for unmarried), insurance type (`1' for private or managed care, and `0' for Medicaid or other types), and income level (categorized as `0' for household incomes below \$30,000, `1' for incomes between \$30,000 and \$75,000, and `2' for incomes above \$75,000). All considered covariates are time-independent variables. 

We evaluated two modifiers: the mother's education level, categorized as ‘0' for individuals without a bachelor's degree, and ‘1' for those with a bachelor's degree or higher, and the mother's pre-pregnancy Body Mass Index (BMI in $\mathrm{kg/m^2}$).  Each modifier was examined independently. The summary statistics for each racial/ethnic group are presented in Table \ref{real:tab:1}.

{
\begin{table}[!htbp] \centering 
  \caption{Final sample characteristics of  the Fetal Growth - Singletons Study} 
  \label{real:tab:1} 
\begin{tabular}{@{\extracolsep{5pt}}lcccc}
\\[-1.8ex]\hline 
\hline \\[-1.8ex] 
& \multicolumn{4}{c}{\% \& Mean (SD)} \\
\makecell[l]{Variables \\} & \makecell{Non-Hispanic \\White  \\ $n=460$ (31\%)} & \makecell{Non-Hispanic \\ Black  \\ $n=359$ (25\%)} & \makecell{Hispanic\\  \\ $n=394$ (27\%)} & \makecell{Non-Hispanic \\ Asian/Pacific Islander \\ $n=251$ (17\%)} \\ 
\hline \\[-1.8ex] 
Pre-pregnancy BMI ($\mathrm{kg/m^2}$) & 23.1 (2.8) & 23.9 (3.2)  & 24.4 (3.1) & 22.2 (2.5) \\ 
Height (cm) & 165.6 (6.7) & 164.8 (6.6) & 160.2 (6.2) & 160.3 (6.1) \\ 
Weight (kg) & 65.7 (9.6) & 67.3 (11.4) & 64.4 (9.8) & 57.0 (8.1) \\ 
Age (years) & 30.3 (4.2) & 25.8 (5.4)  & 27.3 (5.4) & 30.8 (4.2) \\ 
Parity &    \\ 
$~~$More than 0  & 47.0\%  & 53.2\% & 61.4\% & 47.8\%\\
$~~$None  & 53.0\% & 46.8\%  & 38.6\% & 52.2\% \\
Employment &  \\ 
$~~$Yes & 83.3\% & 75.5\%  & 65.2\% & 71.7\%\\ 
$~~$No & 16.7\% & 24.5\%  & 34.8\% & 28.3\% \\ 
Marital Status & \\ 
$~~$Married or living with partner & 95.0\% & 51.0\% & 74.9\% & 92.0\%\\ 
$~~$Not married  & 5.0\% & 49.0\% & 25.1\% & 8.0\%\\ 
Insurance& \\ 
$~~$Private or Managed Care  & 95.4\% & 51.8\% & 43.9\% & 84.1\%\\ 
$~~$Other  & 4.6\% & 48.2\% & 56.1\% & 15.9\%\\ 
Income & \\
$~~$Above \$75,000 & 77.1\% & 23.7\% & 21.1\% & 57.8\%\\ 
$~~$\$30,000 to \$75,000 & 19.2\% & 28.2\% & 40.7\% & 25.5\%\\ 
$~~$Below \$30,000 & 3.7\% & 48.2\% & 38.3\% & 16.7\%\\ 
Education & \\
$~~$Bachelor's degree or higher &  77.0\% & 25.9\% & 19.8\% & 70.9\%\\ 
$~~$Without bachelor's degree & 23.0\% & 74.1\% & 80.2\% & 29.1\%\\ 
\hline
\end{tabular}

\end{table} 
}


\subsection{Disparity decomposition with mother's education as a modifier}\label{sec:mLDDedu}

In this section, we apply the proposed mLDD method to decompose the potential disparity in fetal growth trajectories with the mother's education level, low-educated (`0') vs. high-educated (`1'), as a modifier. For comparative purposes, we also conduct an analysis using the   LDD method.

\begin{figure}[h]
\centering
\subfigure[{NHW vs. NHB}]{\includegraphics[width=1\textwidth, height=0.23\textheight]{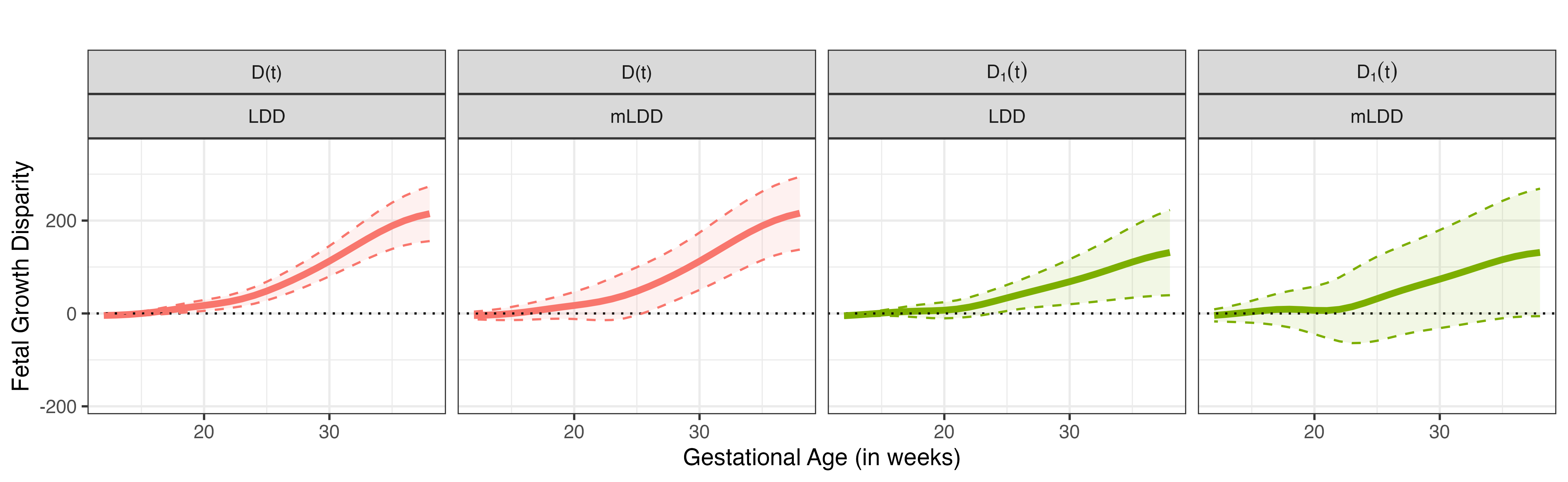}}
\subfigure[NHW vs. HIS]{\includegraphics[width=1\textwidth, height=0.23\textheight]{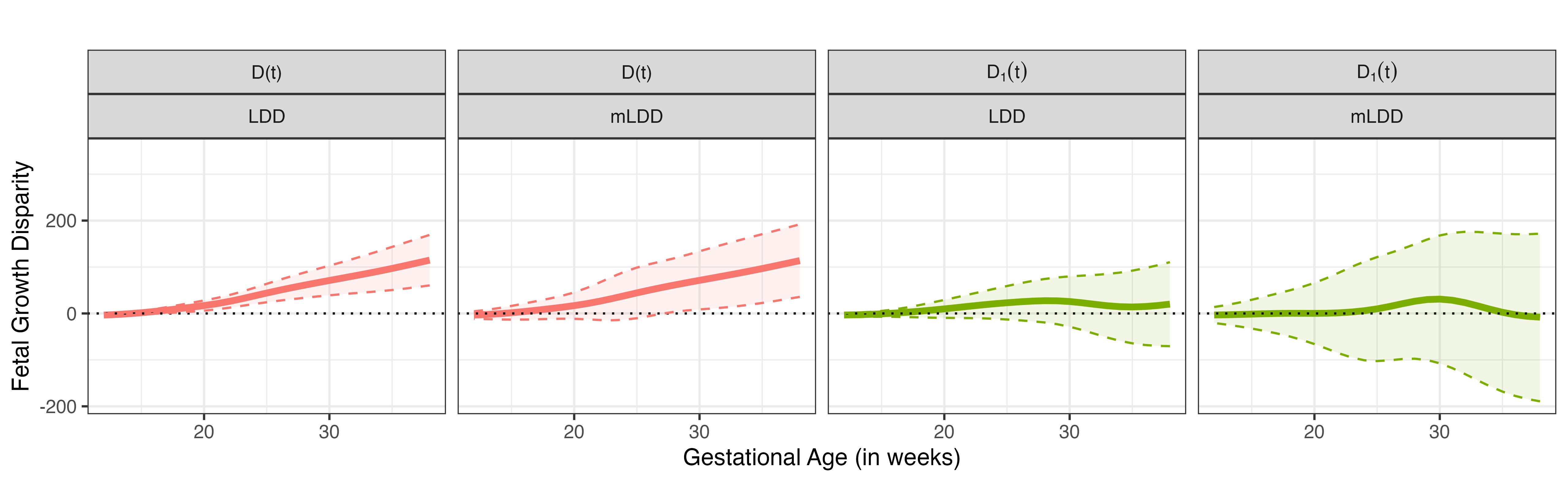}}
\subfigure[NHW vs. API]{\includegraphics[width=1\textwidth, height=0.23\textheight]{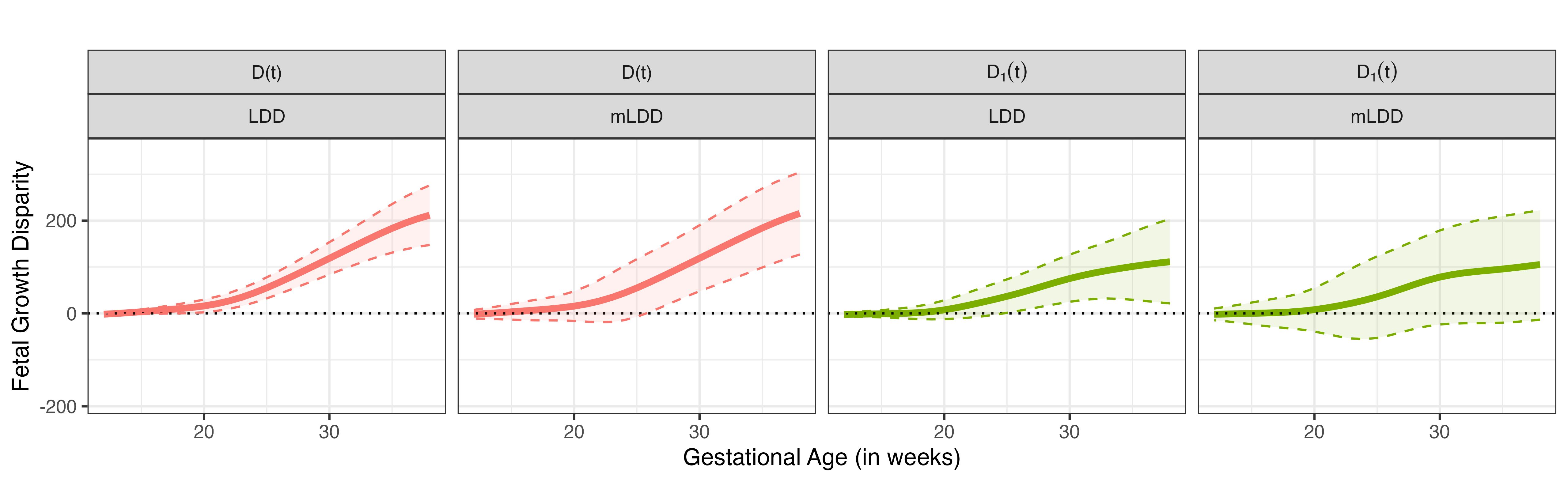}}
\caption{Fetal growth disparity decomposition using LDD and mLDD, considering the mother’s education as a modifier. In panels (a), (b), and (c), NHW is set as the majority group.  In each panel, the first two plots represent the overall disparity, denoted by $D(t)$, and the subsequent two plots depict the unexplained disparity due to the covariates, denoted by   $D_1(t)$. Solid lines represent the estimated disparities and the dashed lines represent the 95\% confidence bands.
}
\label{fig:2}
\end{figure}

In Figure \ref{fig:2}, the first two plots demonstrate the overall disparity and the last two plots show the unexplained disparity as obtained from the LDD and mLDD. Theoretically, the overall disparity obtained via the sum of the unexplained disparity and the explained disparity in  \eqref{eq:naive} and the sum of $D_1(t)$, $D_2(t)$, $D_3(t)$ in  \eqref{eq:1} should be the same. Although it is challenging to directly assess how accurately the proposed mLDD approach can estimate $D_1(t)$, $D_2(t)$, and $D_3(t)$ due to its complex form, we can utilize the result that disparity decomposition obtained from the LDD and mLDD yield the almost identical overall disparity to support the efficacy of the proposed mLDD. Furthermore, the overall disparity reveals that fetuses from NHW mothers exhibit faster growth after week 20 compared to those from mothers of other racial groups.

Additionally, we explore the impact of maternal education level on fetal growth disparity, analyzing it from two distinct viewpoints through the LDD and mLDD approaches. 
To facilitate a clear comparison between LDD and mLDD, we further decompose the explained disparity of LDD in (\ref{eq:naive}) into two components, analogous to the approach used in mLDD:
{
$$
\underbrace{\left[E \{\mathbf{X}^{M}\} - E \{\mathbf{X}^{m}\} \right]\boldsymbol{\beta}^{m}(t)}_\textrm{\small{explained disparity due to $\mathbf{X}$}}+\underbrace{\{E(Z^{M}) - E(Z^{m})\}\beta_{p+1}^{m}(t)}_\textrm{\small{explained disparity due to $Z$}}.
$$
}

\begin{figure}[h]
\centering
\subfigure[NHW vs. NHB]{\includegraphics[width=1\textwidth, height=0.23\textheight]{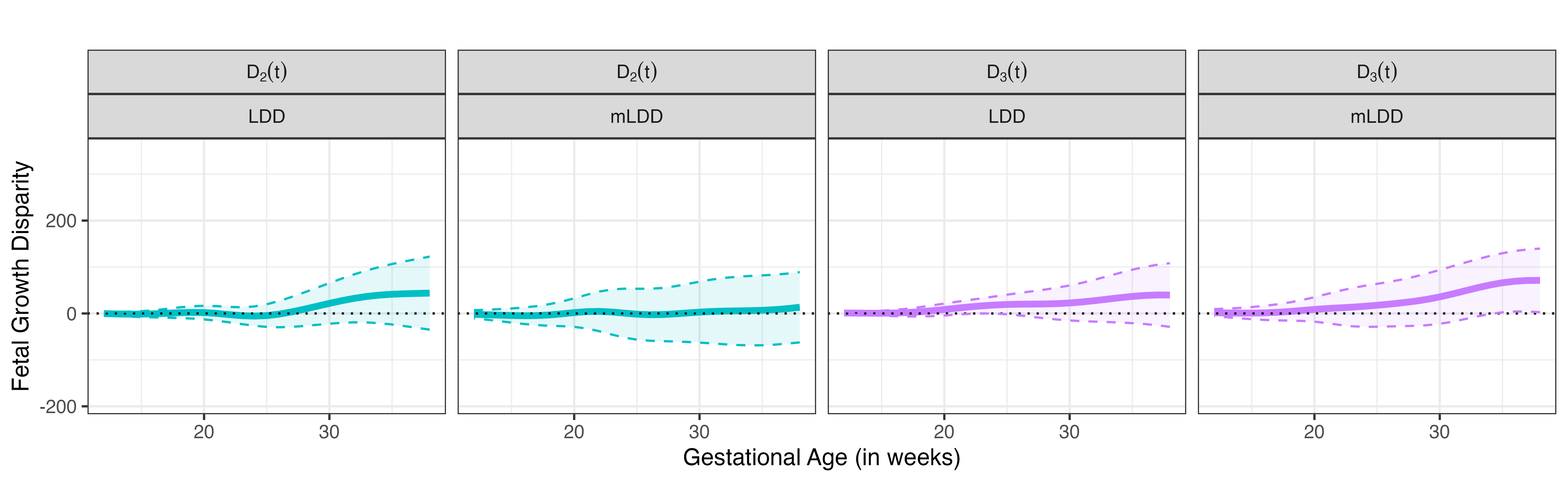}}
\subfigure[NHW vs. HIS]{\includegraphics[width=1\textwidth, height=0.23\textheight]{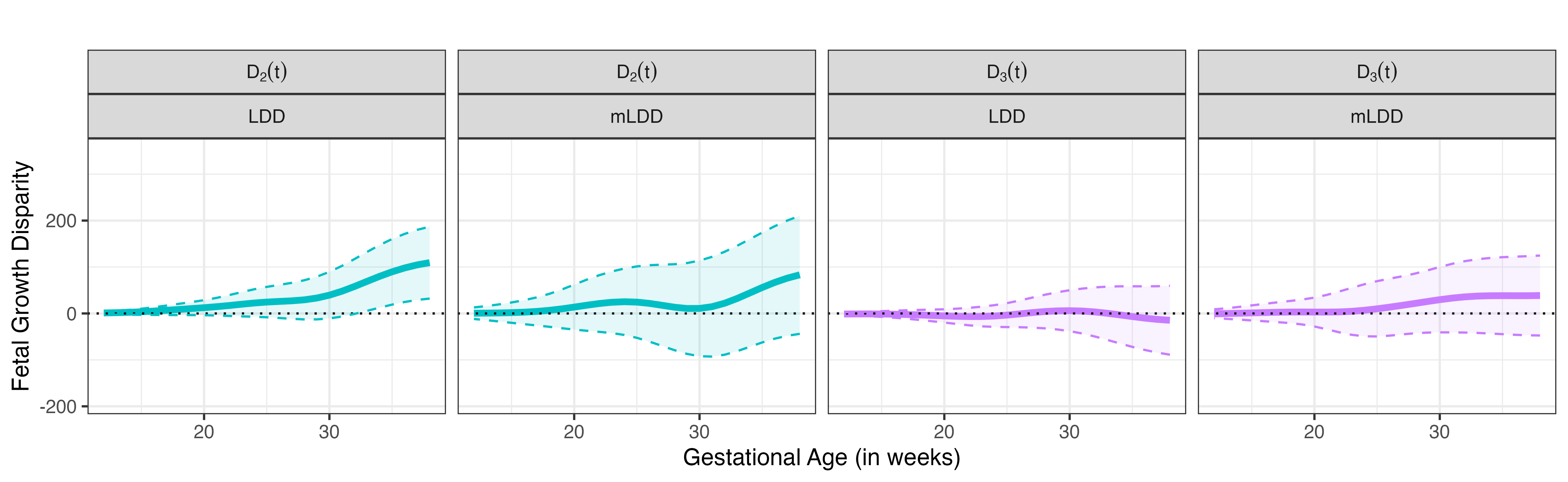}}
\subfigure[NHW vs. API]{\includegraphics[width=1\textwidth, height=0.23\textheight]{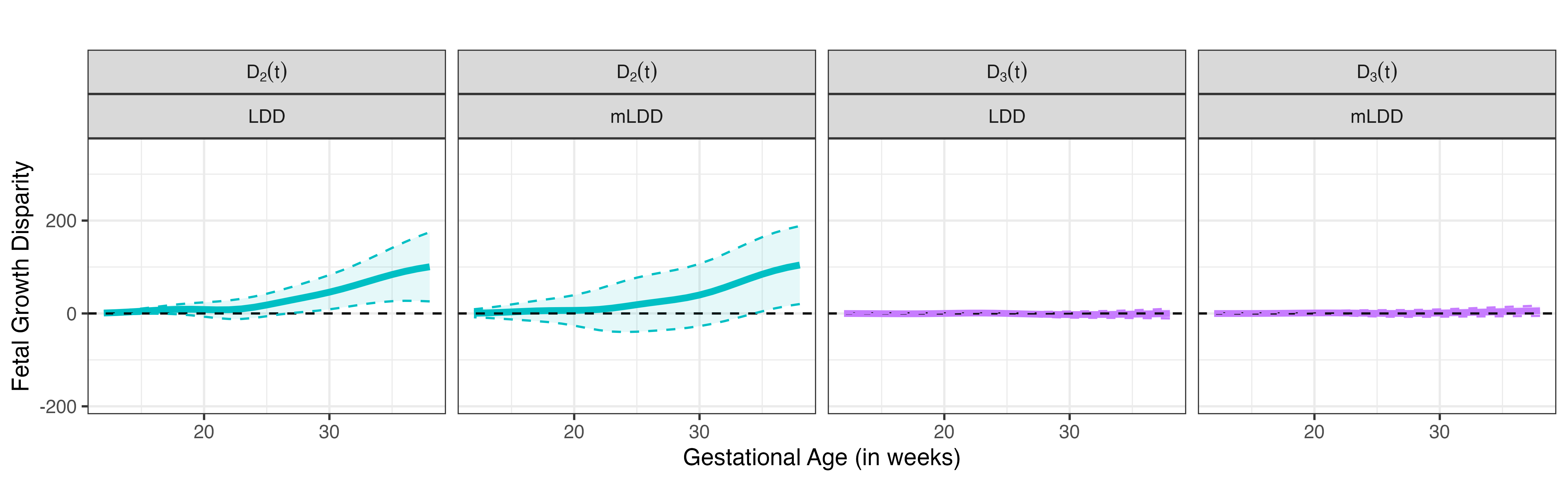}}
\caption{Fetal growth disparity decomposition using LDD and mLDD, considering the mother’s education as a modifier. In panels (a), (b), and (c), NHW is set as the majority group. In each panel, the first two plots represent {explained disparity due to the covariates} except for the modifier, denoted by $D_2(t)$, and  the subsequent two plots depict the explained disparity due to the modifier, denoted by $D_3(t)$. The solid lines represent the estimated disparities and the dashed lines represent the 95\% confidence bands.}
\label{fig:3}
\end{figure}

In Figure \ref{fig:3}, the explained disparities from LDD and mLDD are presented. In Figure  \ref{fig:3}(a), the first two plots present the disparity associated with covariates except the mother's education, but there is a distinction:  the LDD plot considers the mother's education as a covariate, similar to the other covariates
, while the mLDD plot treats the mother's education as a modifier. 
Figure \ref{fig:3}(a) reveals that the explained disparity due to covariates, except the mother's education, remains near zero over time in the mLDD model. However,  although it is not statistically significant, in the LDD model there is a slight amount of explained disparity observed in the later weeks. 
The last two plots of Figure \ref{fig:3}(a) specifically highlight the role of the mother's education ($Z$) as a covariate in LDD and as a modifier in mLDD. It seems that the mother's education amplifies the disparity between  NHW and  NHB over time through its interaction with other covariates.  Similar observations were made in Figure \ref{fig:3}(b) for HIS group.  However, this pattern is not evident for the API group, possibly due to the similar distribution of mothers' education between NHW and API, as indicated in Table \ref{real:tab:1}. Specifically, the proposed mLDD method indicates the explained disparity due to the difference in the distribution of mothers' education level between NHW and NHB mothers becomes barely significant after week 35.
Our findings uncover the role of the mother's education as a modifier, which is not discernible through  LDD. This is attributed to the potential impact of the mother's education on the development of disparities over time and its possible role in the interplay of covariates during the period.

\subsection{Conditional disparity decomposition with mother's education as a modifier}\label{sec:cmLDDedu}

We employ the cmLDD method to examine how the mother's education affects longitudinal disparities in fetal growth, assigning a low education level to the minority group ($z^m=0$) and a high level to the majority group ($z^M=1$). Additionally, we reverse the education levels between the groups, analyzing the situation where the majority has a lower education level ($z^M=0$) and the minority has a higher education level ($z^m=1$). When the education levels are equal across both groups (i.e., $z^M = z^m$),  no disparity is observed due to the modifier, resulting in a disparity measure of zero. This approach allows us to assess the potential decrease in disparity under these circumstances. Unlike mLDD, cmLDD provides the ability to examine disparities at specific modifier levels, offering insights not possible with mLDD.

\begin{figure}[h]
\centering
\subfigure[NHW vs. NHB; ($z^M$ = 1, $z^m$ = 0)]{\includegraphics[width=1\textwidth, height=0.23\textheight]{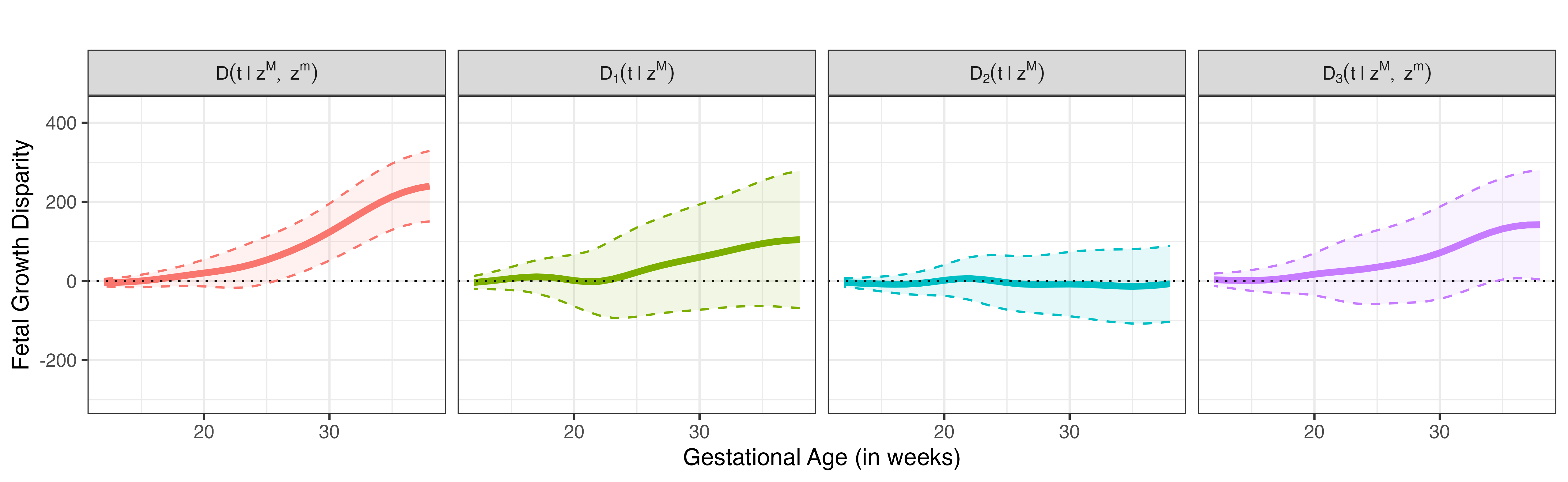}}
\subfigure[NHW vs. HIS; ($z^M$ = 1, $z^m$ = 0)]{\includegraphics[width=1\textwidth, height=0.23\textheight]{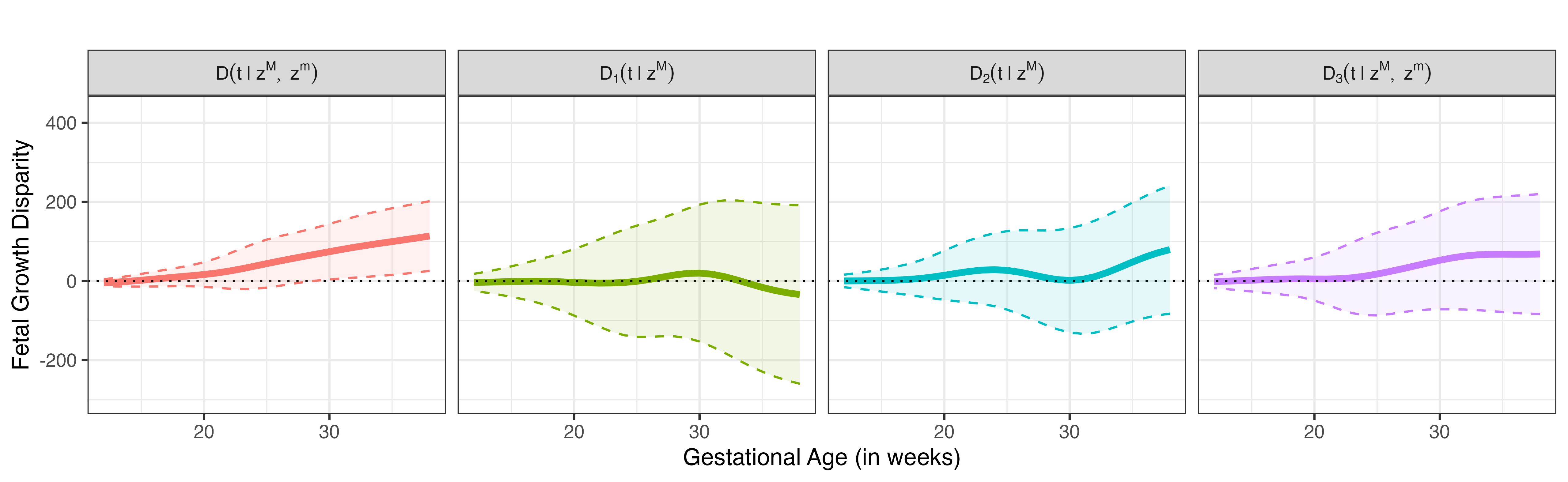}}
\subfigure[NHW vs. API; ($z^M$ = 1, $z^m$ = 0)]{\includegraphics[width=1\textwidth, height=0.23\textheight]{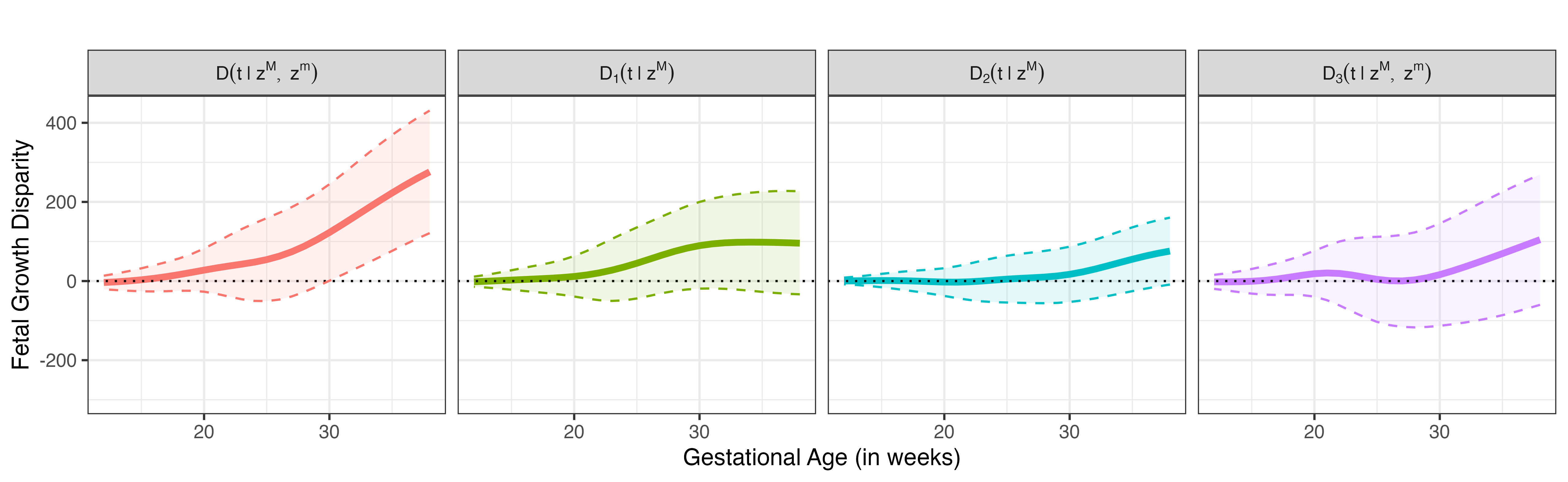}}
\caption{Fetal growth disparity decomposition  with the cmLDD method, using mother’s education level as the modifier where
 $z^M = 1$ and $z^m = 0$.  In panels (a), (b), and (c), NHW is set as the majority group. The solid lines represent the estimated disparities, and the dashed lines represent the 95\% confidence bands.
}
\label{fig:3.1}
\end{figure}

\begin{figure}[h]
\centering
\subfigure[NHW vs. NHB; ($z^M$ = 0, $z^m$ = 1)]{\includegraphics[width=1\textwidth, height=0.23\textheight]{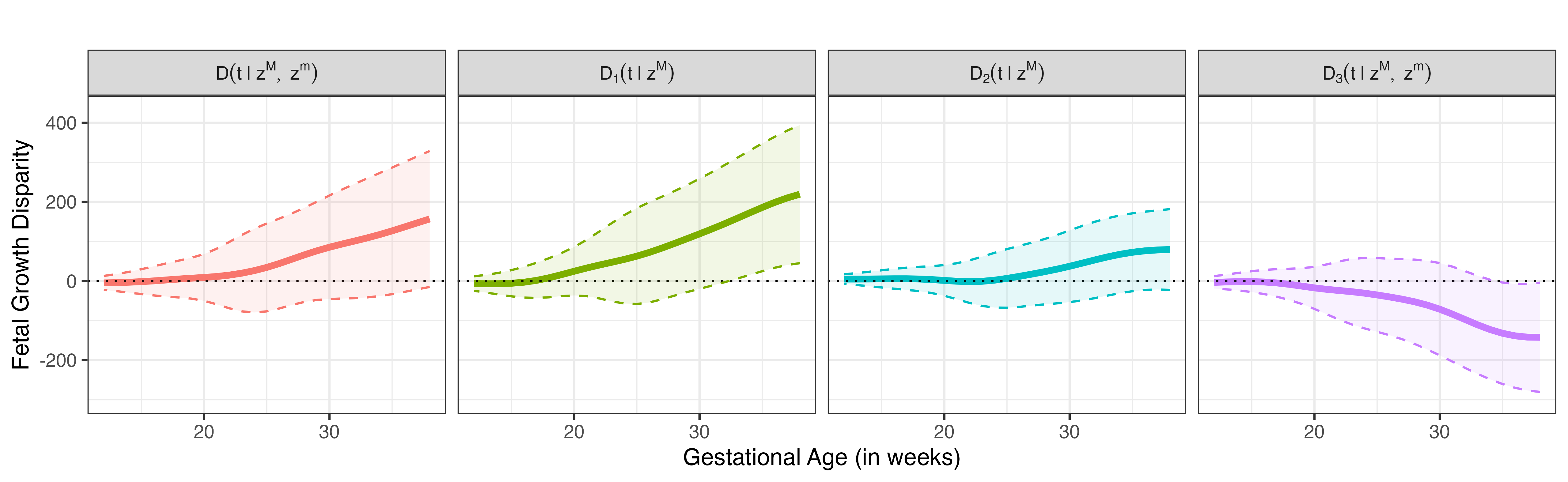}}
\subfigure[NHW vs. HIS; ($z^M$ = 0, $z^m$ = 1)]{\includegraphics[width=1\textwidth, height=0.23\textheight]{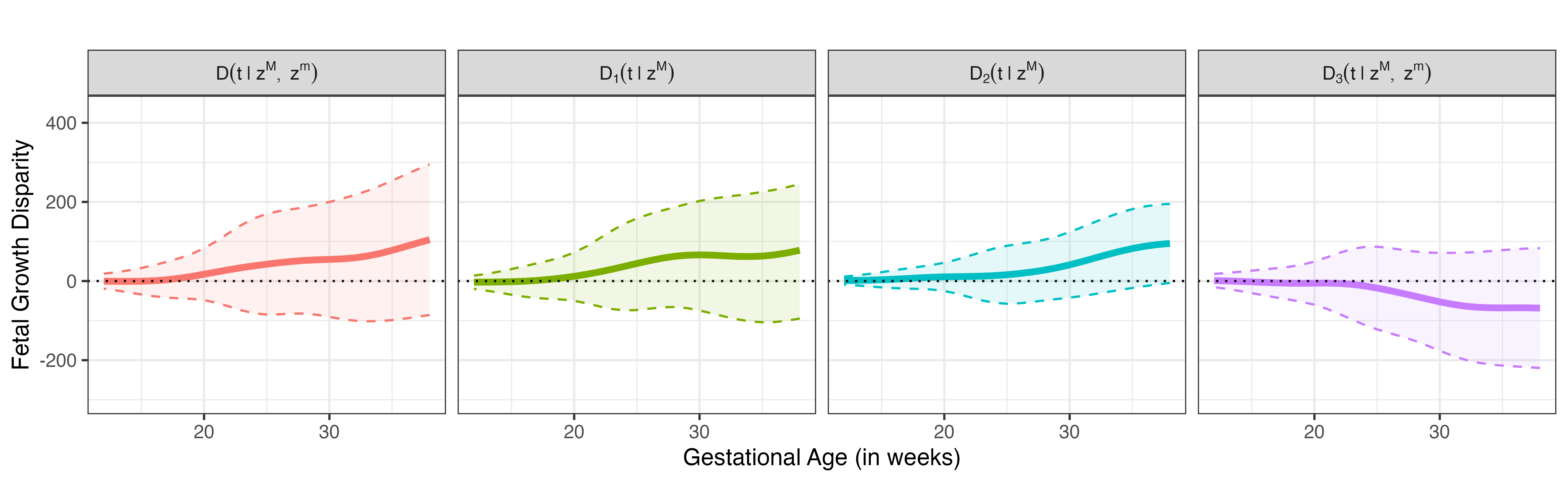}}
\subfigure[NHW vs. API; ($z^M$ = 0, $z^m$ = 1)]{\includegraphics[width=1\textwidth, height=0.23\textheight]{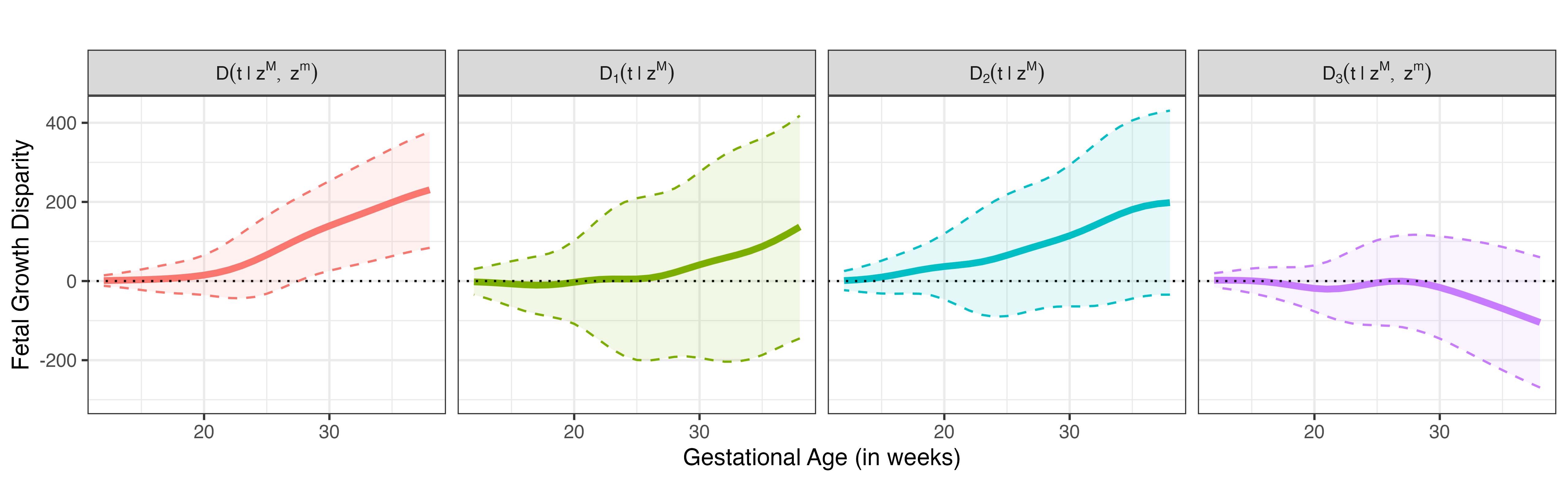}}
\caption{Fetal growth disparity decomposition  with the cmLDD method, using mother’s education level as the modifier where
 $z^M = 0$ and $z^m = 1$.  In panels (a), (b), and (c), NHW is set as the majority group. The solid lines represent the estimated disparities, and the dashed lines represent the 95\% confidence bands.
}
\label{fig:3.2}
\end{figure}

Figures \ref{fig:3.1}-\ref{fig:3.2} displays the estimated $D(t|z^M,z^m), D_1(t|z^M), D_2(t|z^M)$ and $D_3(t|z^M,z^m)$ from cmLDD, representing the overall disparity, unexplained disparity,  explained disparity due to {covariates} at specific modifier levels, and explained disparity due to the modifier levels, respectively.  

A key observation from Figure \ref{fig:3.1}  is the more pronounced influence of the mother's education in the cmLDD analysis compared to mLDD, highlighting the difference in their analytical frameworks.
The cmLDD model's $D_3(t|z^M=1,z^m=0)$  indicates the potential decrease in disparity that could be achieved if the education level of mothers in the minority group,  currently lower, were elevated to match that of the majority group.  
In contrast to the patterns observed in {Figure}   \ref{fig:3}, $D_3(t|z^M=1,z^m=0)$ in Figure \ref{fig:3.1} (a) suggests a significant opportunity for reducing disparity among   NHB mothers with lower education levels. Additionally, there is a noticeable potential for reduction in disparity for the API group, as indicated by   $D_3(t|z^M=1,z^m=0)$  in Figure \ref{fig:3.1} (c). Although this increase is not statistically significant, it is a distinction not visible in  Figure \ref{fig:3}.

On the other hand, Figure \ref{fig:3.2}  presents the cmLDD results where the majority group (NHW) has lower education ($z^M=0$) and the minority group has a high education ($z^m=1$).  In Figure \ref{fig:3.2},   we observe smaller overall fetal growth disparities, denoted as $D(t|z^M=0,z^m=1)$, between NHW and NHB,  NHW and HIS, and  NHW and API,  compared to the disparities observed as $D(t|z^M=1,z^m=0)$ in Figure \ref{fig:3.1}. Notably, in Figure \ref{fig:3.2}, the explained disparity due to the modifier $D_3(t|z^M=0,z^m=1)$ trends negative over time, in contrast to  Figure \ref{fig:3.1}, where $D_3(t|z^M=1,z^m=0)$ trends positive. This illustrates the significant role of mothers' education in fetal growth disparities.

\subsection{Decomposition of disparity with mother’s BMI as a modifier}

\begin{figure}[h]
\centering
{\small (A) NHW vs. NHB}\\
\subfigure[($z^M$=22.7, $z^m$=19.8)]{\includegraphics[width=0.25\textwidth, height=0.25\textheight]{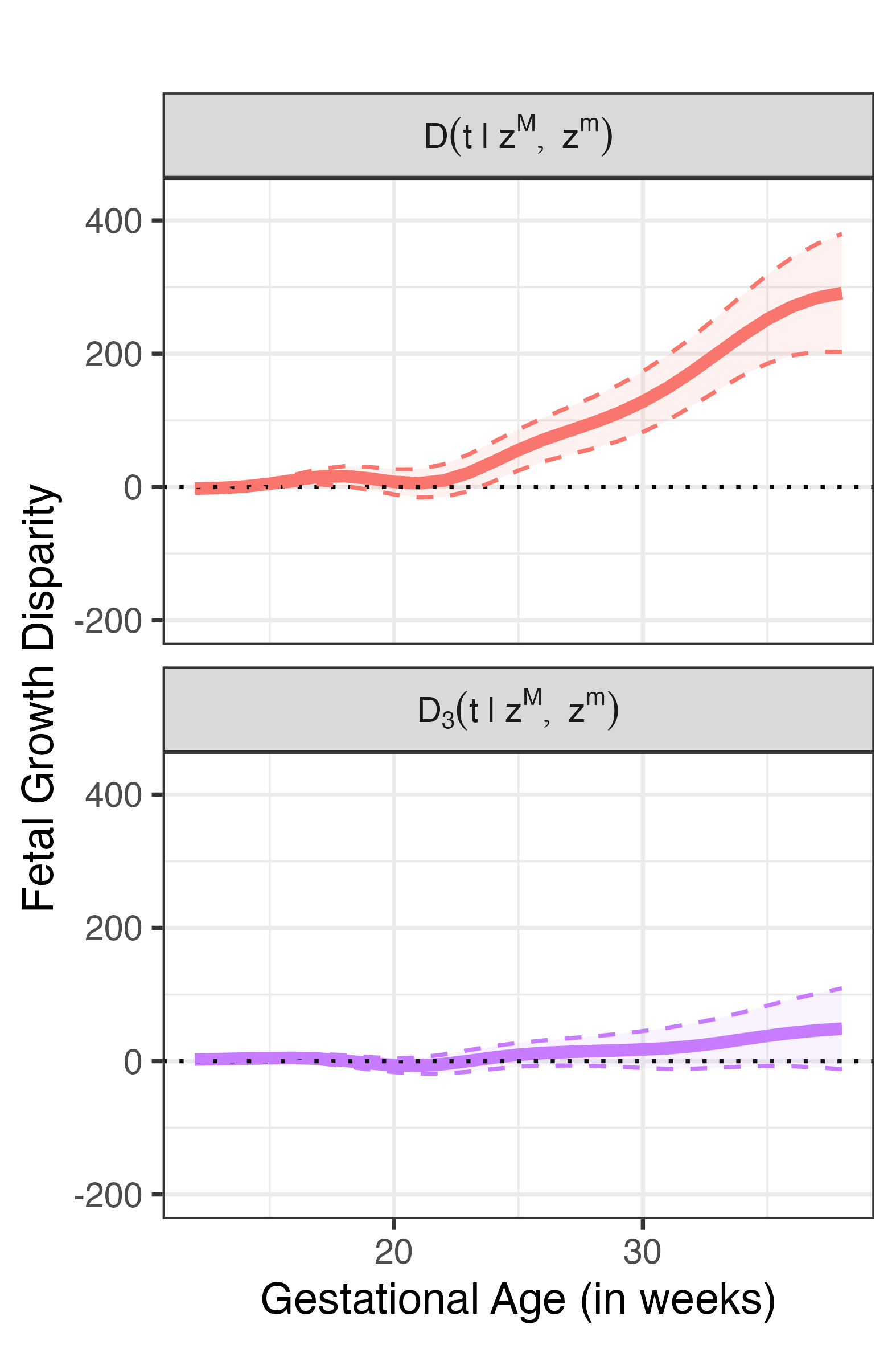}}
\subfigure[($z^M$=22.7, $z^m$= 23.9)]{\includegraphics[width=0.25\textwidth, height=0.25\textheight]{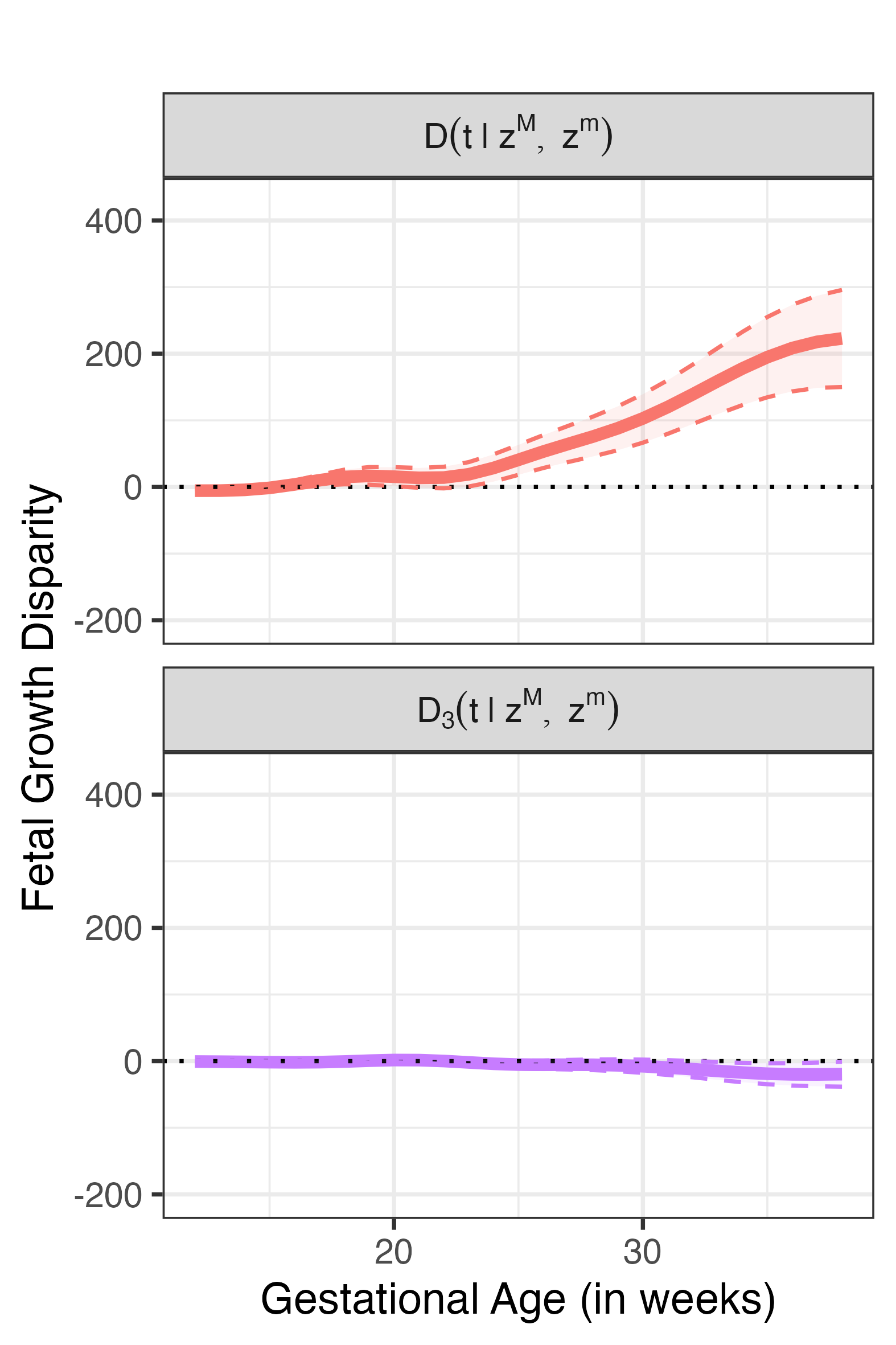}}
\subfigure[($z^M$=22.7, $z^m$=28.6)]{\includegraphics[width=0.25\textwidth, height=0.25\textheight]{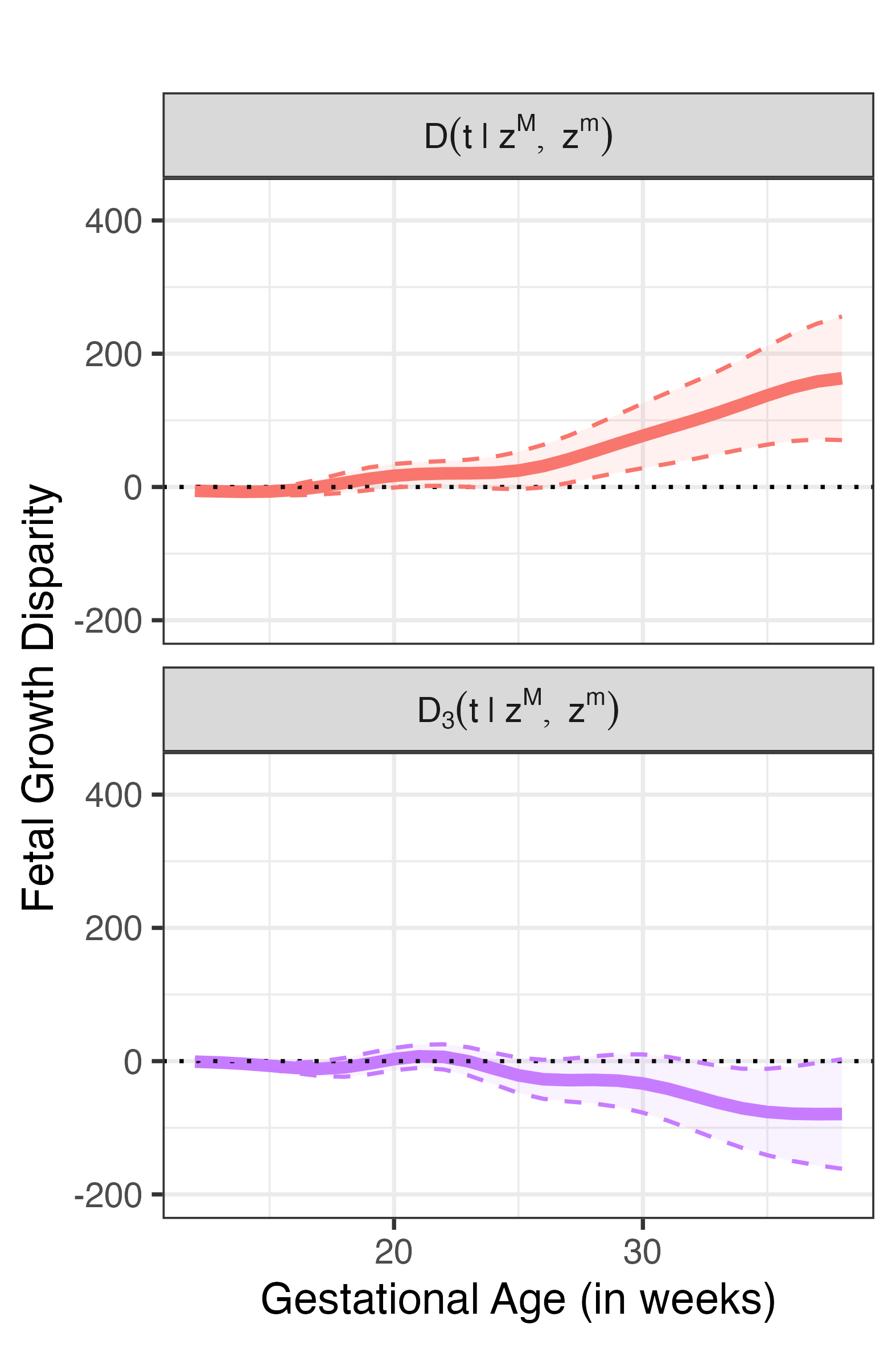}}\\ 
\medskip
{\small (B) NHW vs. HIS}\\
\subfigure[($z^M$=22.7, $z^m$=20.5)]{\includegraphics[width=0.25\textwidth, height=0.25\textheight]{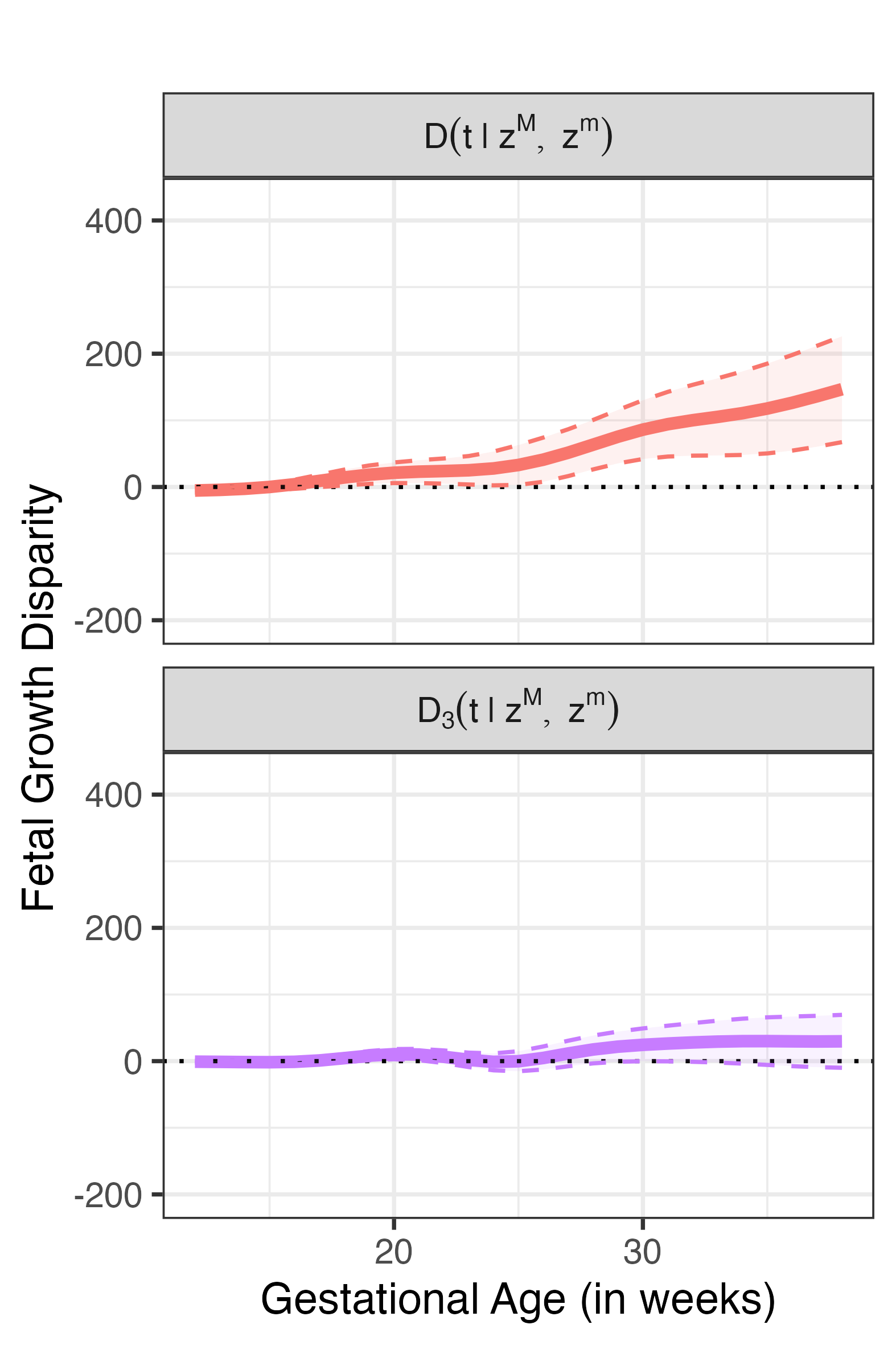}}
\subfigure[($z^M$=22.7, $z^m$= 24.1)]{\includegraphics[width=0.25\textwidth, height=0.25\textheight]{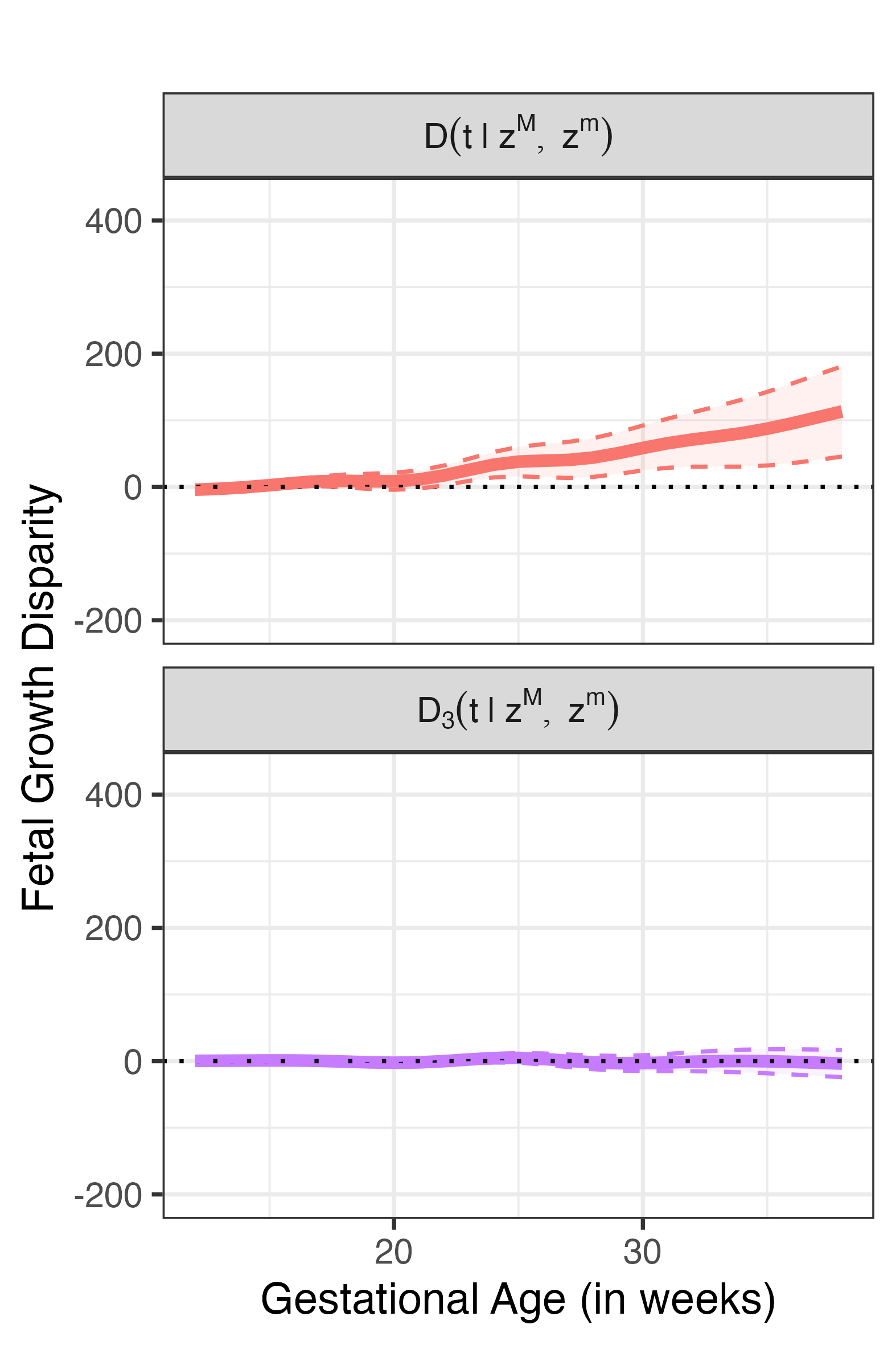}}
 \subfigure[($z^M$=22.7, $z^m$= 28.0)]{\includegraphics[width=0.25\textwidth, height=0.25\textheight]{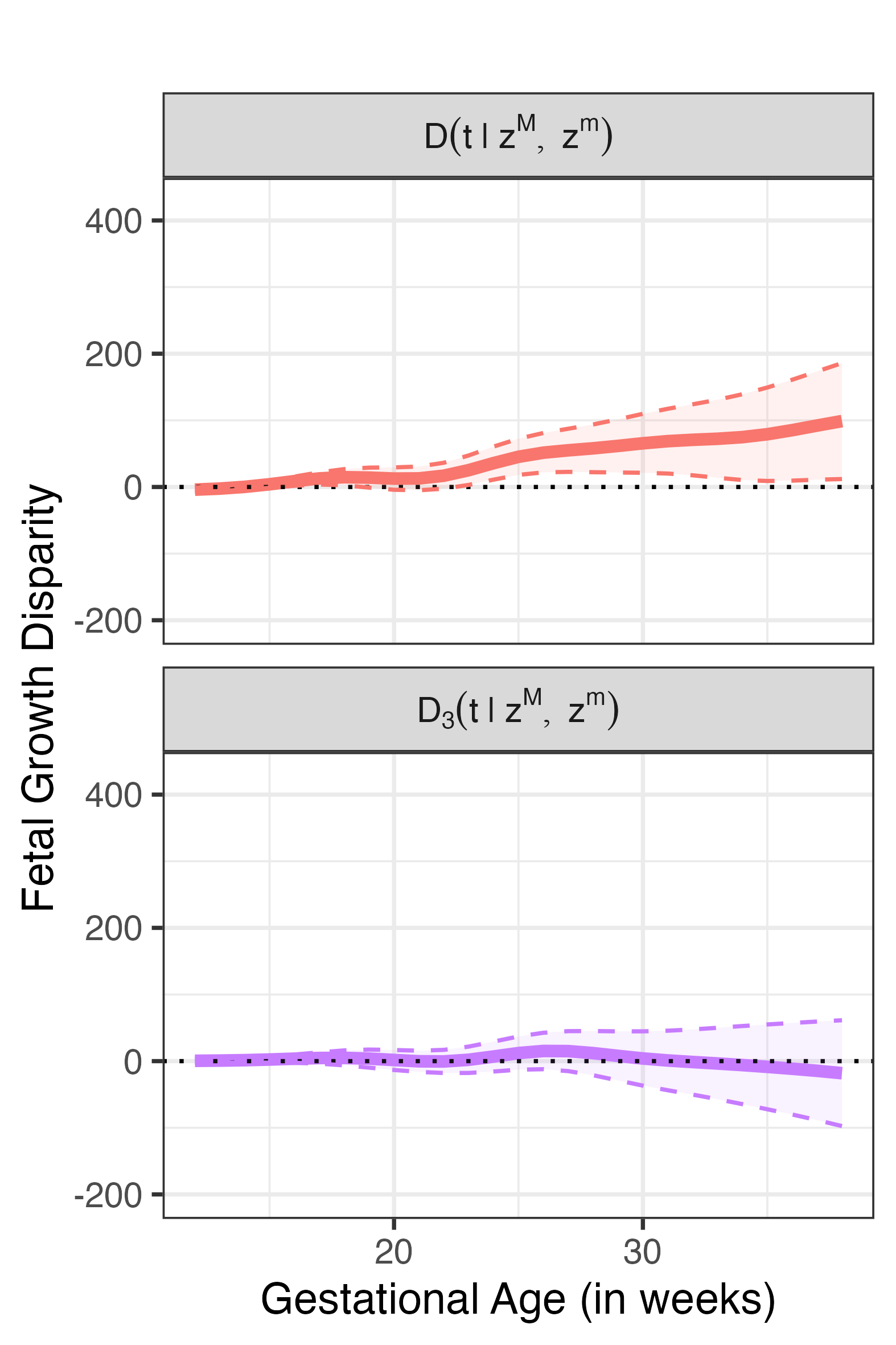}}\\ 
 \medskip
{\small (C) NHW vs. API}\\
\subfigure[($z^M$=22.7,  $z^m$=19.3)]{\includegraphics[width=0.25\textwidth, height=0.25\textheight]{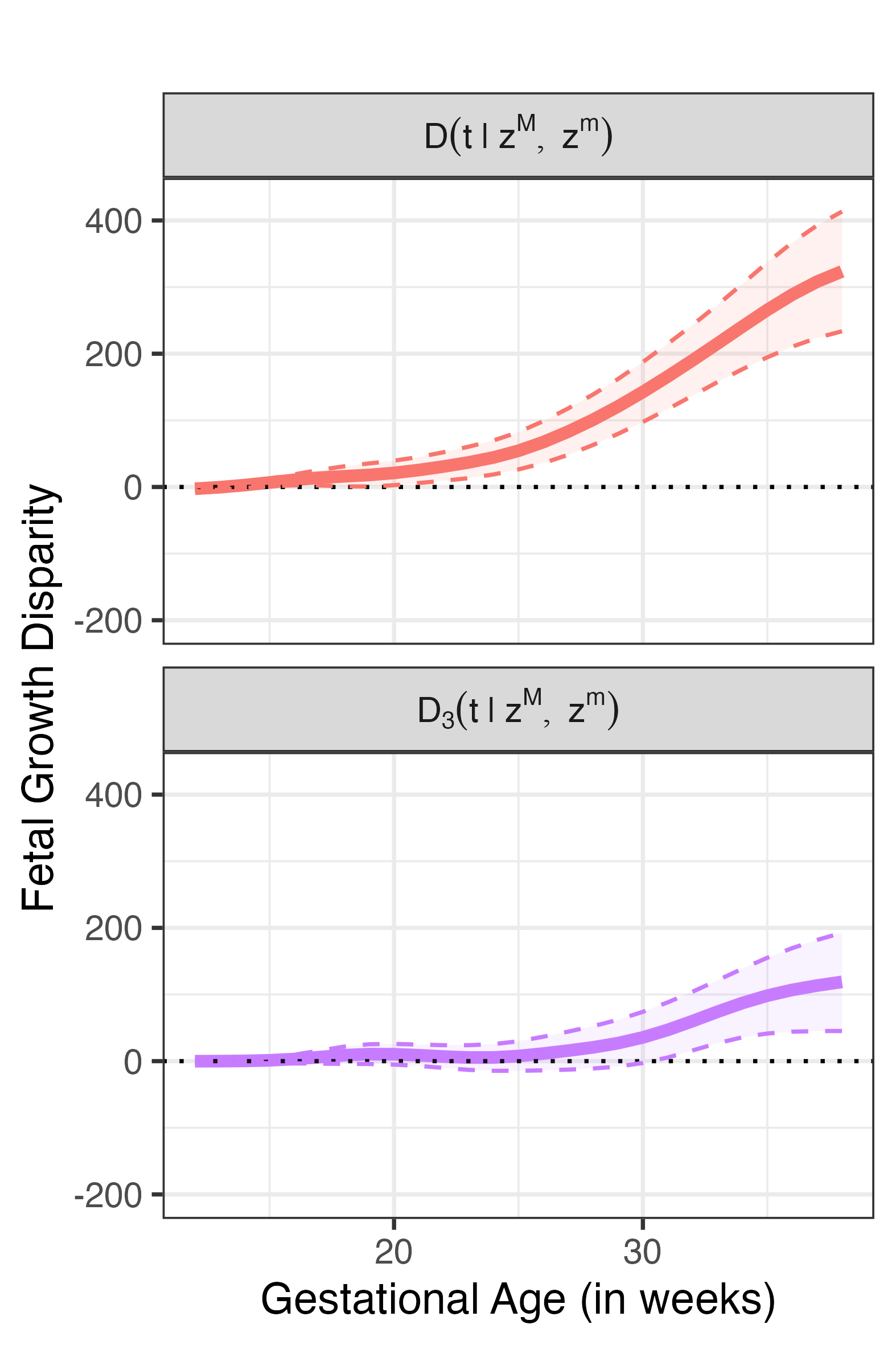}}
\subfigure[($z^M$=22.7, $z^m$= 21.5)]{\includegraphics[width=0.25\textwidth, height=0.25\textheight]{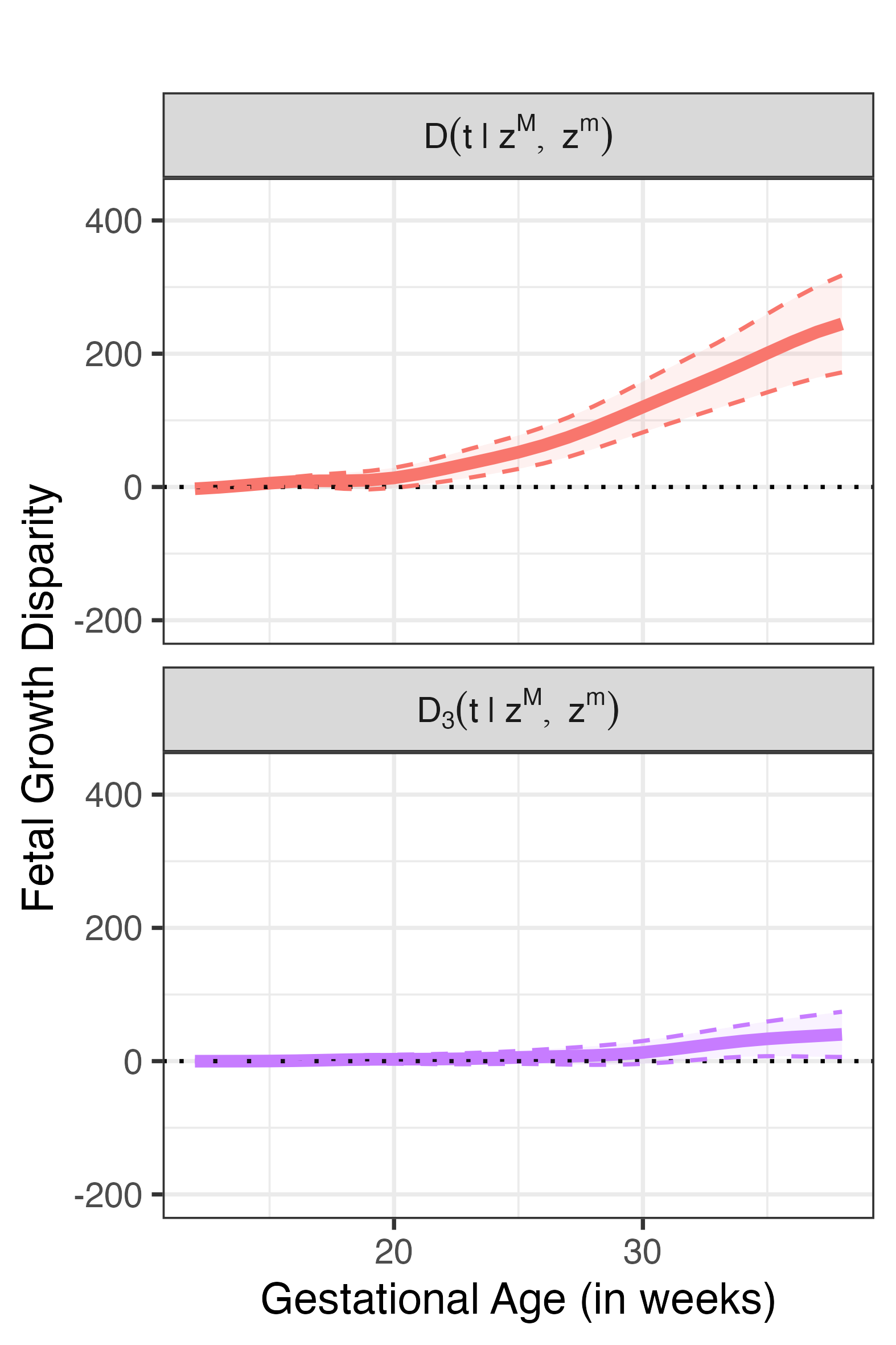}}
\subfigure[($z^M$=22.7,  $z^m$=25.7)]{\includegraphics[width=0.25\textwidth, height=0.25\textheight]{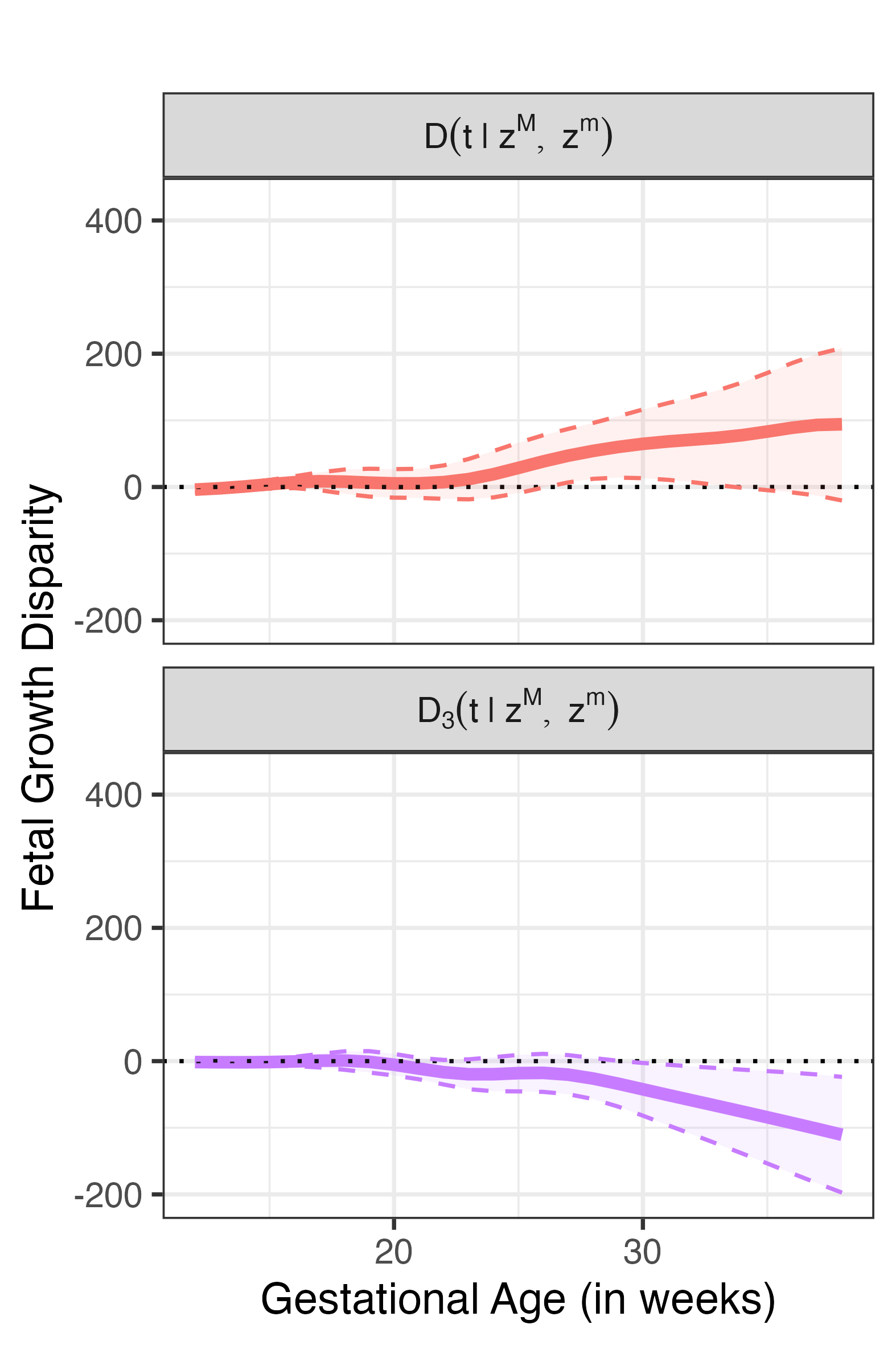}}
\caption{Fetal growth disparity decomposition with BMI as modifier through cmLDD, where the majority group is NHW. The analysis considers modifier levels at the NHW mother's median BMI ($z^M=22.7$) and the minority groups's mothers' BMI at the 10th, 50th, and 90th percentiles. The solid lines represent the estimated disparities and the dashed lines represent the 95\% confidence bands.}
\label{fig:4}
\end{figure}

We illustrate the application of the cmLDD, where the modifier is a continuous variable, using the mother's BMI as an example. To investigate the influence of  mothers' BMI on the disparity in fetal growth between  majority and minority groups, we set the value of $z$ as follows:
For the majority group (NHW), we set $z^M=22.7$, which is the sample median BMI among NHW mothers. For the minority groups, we selected $z^m$ to represent the 10th, 50th, and 90th BMI quantiles. This choice is driven by our sample consisting of non-obese women, for whom the  BMI range is relatively narrow, and the differences between the first and third quartiles are slight. The cmLDD decomposition in (\ref{eq:cmLDD}) is $D(t|z^M,z^m)=D_1(t|z^M)+D_2(t|z^M)+D_3(t|z^M,z^m)$, where $z^m$ is a BMI quantile of interest for minority group. 
Given that $D_1(t|z^M)$ and $D_2(t|z^M)$ do not depend on $z^m$, 
we only present the overall disparity, $D(t|z^M,z^m)$, and the disparity explained by  the modifier value, $D_3(t|z^M,z^m)$.

Figure \ref{fig:4} illustrates the fetal growth disparity between the majority group (NHW) and minority groups, using specific maternal BMI values as benchmarks.   We used   $z^M=22.7$ for NHW, representing the median pre-pregnancy BMI.  For NHB, we set  $z^m=19.8$,  $z^m= 23.9$, and $z^m=28.6$ to reflect the 10th, 50th, and 90th percentiles, respectively. Similarly, for HIS, the values are  $z^m=20.5$, $z^m= 24.1$ and  $z^m= 28.0$; and for API, $z^m=19.3$, $z^m= 21.5$, and  $z^m=25.7$, corresponding to the same percentiles.  
In each of the sub-figures (a), (b), and (c) of Figure \ref{fig:4}, the top figures illustrate the overall disparity at time \(t\), denoted as \(D(t|z^M,z^m)\), and  the bottom figures depict the disparity attributable to the modifiers \(z^M\) and \(z^m\) at time \(t\), specifically \(D_3(t|z^M,z^m)\).

The findings presented in  Figure \ref{fig:4}(A) reveal notable overall disparities in fetal growth emerging after week 30  between fetuses of NHW mothers at the median BMI and those of NHB mothers at various BMI quantiles.  Moreover, disparities attributable to BMI differences, particularly those between NHW mothers with median BMI and NHB mothers at the 10th BMI percentile, seem to amplify the overall disparity.  Conversely, NHB mothers at the 90th BMI percentile show a trend toward reducing the overall fetal growth disparity in later gestational weeks, though these findings are not statistically significant. This observation could be attributed to our dataset being comprised exclusively of non-obese women.

 Figure \ref{fig:4}(B) indicates that the overall fetal growth disparities between NHW and HIS groups are less pronounced compared to disparities observed with other racial groups. However, similar patterns of disparity related to the mother's BMI are noted between NHW and HIS, consistent with those observed between NHW and NHB, as we varied  $z^m$ across the  10th, 50th, and 90th percentiles.

In Figure \ref{fig:4}(C), it is evident that the overall fetal growth disparities between NHW mothers at median BMI and the API group across different BMI quantiles significantly diminish as the BMI quantiles of the API group increase. This indicates that the BMI of API mothers has a notable impact on fetal growth, potentially affecting the disparity either directly or indirectly through interactions with other covariates.

\section{Discussion}

In this paper, we introduce a new method for decomposing longitudinal disparity. This approach employs a dynamic model with time-varying coefficients, enabling precise tracking of disparity progression over time. Longitudinal datasets that can track how disparities evolve over time are becoming more prevalent.  For instance, our method is applicable in areas like the analysis of longitudinal disparities in prostate-specific antigen (PSA) levels within prostate cancer research or in cancer antigen-125 (CA-125) levels in ovarian cancer studies across different racial/ethnic groups.


We acknowledge that our results may not support causal interpretations, as the method does not incorporate a formal causal framework. Incorporating such a framework necessitates a detailed understanding of the causal relationships between specific variables and other covariates not only cross-sectionally but also over time to address potential confounding factors explicitly. This knowledge, however, is often not readily available and is supplanted by assumptions. Whereas the desired causality follows only if such assumptions are collectively valid as a whole, their empirical examination is often not feasible and bias arising from the violation of the assumptions is a serious concern. On the other hand, sensitivity analyses may also rely on the speculative causal relationship as they tend to question one part of the causal relationship while relying on the rest. The proposed method sidesteps the highly speculative nature of formal causal inference.

Since we utilized simultaneous confidence bands (SCBs) for inference, this approach led to relatively large confidence bands. For a more practical and focused analysis, we may employ piecewise confidence intervals on specific gestational periods of interest to practitioners.

Although our proposed methods were initially designed for longitudinal data analysis, they are also applicable to cross-sectional data, as described by 
$
Y=\beta_0(Z)+\beta_1(Z)X_1+\cdots+\beta_p(Z)X_p+\sigma(Z)\epsilon.
$
In this context, both marginal and conditional disparity decomposition techniques can be adapted, whether or not the modifier $Z$ is considered. 

Although our proposed longitudinal disparity decomposition method is demonstrated using fetal growth data from singletons, this dataset may not fully reflect disparities between majority and minority groups. The NICHD fetal growth dataset is centered around non-obese women, indicating that careful consideration is needed when interpreting our results. Furthermore, fetal growth unfolds over a relatively short timeframe. The impact of modifiers like the mother's education and BMI on fetal growth disparities may be modest, as suggested in our analysis \citep{lambert2020maternal}.

Despite the outlined limitations, our proposed model for longitudinal disparity decomposition successfully highlighted the significant impact of mother's education and BMI as modifiers on fetal growth disparities - a dimension often elusive with alternative methodologies \citep{GARDOSI2009, lambert2020maternal}. This underlines the value of our approach in elucidating the complex trajectories of longitudinal disparities, reinforcing its importance in advancing our understanding of these issues.


\begin{acks}[Acknowledgments]

 The NICHD Fetal Growth Studies - Singletons was supported by the Division of Population Health, Division of Intramural Research, \textit{Eunice Kennedy Shriver} National Institute of Child Health and Human Development (NICHD), National Institutes of Health for the NICHD Fetal Growth Studies (Contract Numbers: HHSN275200800013C; HHSN275200800002I; HHSN27500006; HHSN275200800003IC; HHSN275200800014C; HHSN275200800012C; HHSN27520080
0028C; HHSN275201000009C). The authors also thank Dr. Barry Graubard for his valuable discussions.

\end{acks}
\begin{funding}
The first, fourth, and last authors received support from NIH Intramural Research Program. The third author's work was partly funded by NIH Grant P30CA082103. 
\end{funding}
\begin{supplement}
\stitle{Supplement A}
\sdescription{Supplement A outlines the assumptions and the definitions pertinent to Theorem \ref{thm:1}.}
\end{supplement}
\clearpage
\bibliographystyle{imsart-nameyear} 
\bibliography{reference.bib}       

\end{document}


\begin{frontmatter}
\title{Supplemental Material to ``Decomposition of the Longitudinal Disparities: an Application to the Fetal Growth-Singletons Study''}
\runtitle{Decomposition of the Longitudinal Disparities Supplement}

\begin{aug}
\author[A, B]{\fnms{Sang Kyu}~\snm{Lee}\ead[label=e1]{sangkyu.lee@nih.gov}}\footnote[1]{Authors equally contributed},
\author[C]{\fnms{Seonjin}~\snm{Kim}\ead[label=e2]{kims20@miamioh.edu}}$^*$,
\author[D]{\fnms{Mi-Ok}~\snm{Kim}\ead[label=e3]{miok.kim@ucsf.edu}}
\author[E]{\fnms{Katherine L.}~\snm{Grantz}\ead[label=e4]{katherine.grantz@nih.gov}}
\author[B]{\fnms{Hyokyoung G.}~\snm{Hong}\ead[label=e5]{grace.hong@nih.gov}}
\address[A]{Department of Statistics and Probability,
Michigan State University}
\address[B]{Biostatistics Branch, Division of Cancer Epidemiology and Genetics,
National Cancer Institute\printead[presep={,\ }]{e1,e5}}
\address[C]{Department of Statistics, Miami University\printead[presep={,\ }]{e2}}
\address[D]{Department of Epidemiology and Biostatistics, University of California San Francisco\printead[presep={,\ }]{e3}}
\address[E]{Epidemiology Branch, Division of Population Health Research, National Institute of Child Health and Human Development\printead[presep={,\ }]{e4}}
\end{aug}

\end{frontmatter}

\begin{supplement}
\section*{Assumptions for Theorem 3.1}
We define $\bfbeta'_t(t,z)=\partial \bfbeta(t,z)/\partial t$,  $\bfbeta''_t(t,z)=\partial^2 \bfbeta(t,z)/\partial t^2$ $\bfbeta'_z(t,z)=\partial \bfbeta(t,z)/\partial z$, $\bfbeta''_z(t,z)=\partial^2 \bfbeta(t,z)/\partial z^2$,  $f'_t(t,z)=\partial f_{T,Z}(t,z)/\partial t$ and  $f'_z(t,z)=\partial f_{T,Z}(t,z)/\partial z$. We present the following set of assumptions to establish the theoretical properties.
\begin{itemize}
    \item[(A1)] The covariates $\bfX(t)$ are bounded, $E\{\bfX(t)\}=0$ and $\Gamma_X(t)$ is positive-definite and differentiable. In practice,  it is not necessary that covariates are centralized.

\item[(A2)] The kernel function $K(\cdot)$ is symmetric, bounded support and bounded derivative. 

\item[] \textbf{- For continuous $Z$,}

\item[(A3)] $E\{\sigma^4(t,z)\epsilon^4\}$ is continuously differentiable   in a neighborhood of $t$ and $z$ and $E\{\sigma(t,z)^4\epsilon^4\}<\infty$.

\item[(A4)]  $\bfbeta(\cdot,\cdot)$ are four times continuously differentiable  and $\sigma^2(\cdot,\cdot)$ is continuously differentiable in a neighbourhood of $t$ and $z$. $f_{T,Z}(\cdot,\cdot)$  is twice  continuously differentiable  in a neighbourhood of $t$ and $z$ and $f_{T,Z}(t,z)>0$.

\item[] \textbf{- For discrete $Z$,}

\item[(A3)] For any $z$, $E\{\sigma^4(t,z)\epsilon^4\}$ is continuously  differentiable  in a neighborhood of $t$ and $E\{\sigma(t,z)^4\epsilon^4\}<\infty$.

\item[(A4)] For any $z$, $\bfbeta(\cdot,z)$ are four times continuously differentiable  and $\sigma^2(\cdot,z)$ is continuously differentiable in a neighbourhood of $t$. $f_{T}(\cdot)$  is twice  continuously differentiable  in a neighbourhood of $t$ and $f_T(t)>0$.
\end{itemize}
Assumptions (A1)-(A4) are standard assumptions in kernel-based regression. The proof of Theorem 3.1 follows a conventional approach, and therefore, we omit it in this context. Interested readers are encouraged to refer to \citep{wu2000kernel, kim2013unified} for detailed discussions and complete proof.

\end{supplement}


\bibliographystyle{imsart-nameyear} 
\bibliography{reference.bib}       